\documentclass[12pt]{amsart}
\usepackage{latexsym, amsmath, amssymb,amsthm,amsopn,amsfonts,mathrsfs,tikz}
\usepackage{stmaryrd}
\usepackage{version}
\usepackage{epsfig,graphics,color,graphicx,graphpap,dsfont,eufrak}
\usepackage{amssymb}
\usepackage{geometry}
\usepackage{tikz}
\usetikzlibrary{snakes}

\usepackage{graphicx}
\usepackage{subfigure}

\ifx\pdfoutput\undefined
  \DeclareGraphicsExtensions{.eps}
\else
  \ifx\pdfoutput\relax
    \DeclareGraphicsExtensions{.eps}
  \else
    \ifnum\pdfoutput>0
      \DeclareGraphicsExtensions{.pdf}
    \else
      \DeclareGraphicsExtensions{.eps}
    \fi
  \fi
\fi

\newcommand{\bse}{\begin{subequations}}
\newcommand{\ese}{\end{subequations}}

\DeclareFontFamily{OT1}{pzc}{}
\DeclareFontShape{OT1}{pzc}{m}{it}{<-> s * [1.10] pzcmi7t}{}
\DeclareMathAlphabet{\mathpzc}{OT1}{pzc}{m}{it}

\theoremstyle{plain}

\theoremstyle{definition}

\numberwithin{equation}{section}

\def\squarebox#1{\hbox to #1{\hfill\vbox to #1{\vfill}}}

\usepackage{amsxtra}

\usepackage{hyperref}

\newcommand{\e}{\epsilon}

\newcommand{\ra}{\rightarrow}
\newcommand{\al}{\alpha}
\newcommand{\be}{\begin{equation}}
\newcommand{\ee}{\end{equation}}

\newcommand{\om}{\omega}

\begin{document}

\title[The PAT model of population dynamics]{The PAT model of population dynamics}

\author{Z. C. Feng}
\address{Department of Mechanical and Aerospace Engineering, 
University of Missouri, Columbia, MO 65211}
\email{fengf@missouri.edu}

\author{Y. Charles Li}
\address{Y. Charles Li, Department of Mathematics, University of Missouri, 
Columbia, MO 65211, USA}
\email{liyan@missouri.edu}
\urladdr{https://liyan.mufaculty.umsystem.edu}

\begin{abstract}
We introduce a population-age-time (PAT) model which describes the temporal evolution of the population distribution in age.
The surprising result is that the qualitative nature of the population distribution dynamics is robust with respect to the birth 
rate and death rate distributions in age, and initial conditions. When the number of children born per woman is $2$, the population distribution approaches an asymptotically steady state of a kink shape; thus the total population approaches a constant.
When the number of children born per woman is greater than  $2$, the total population increases without bound; and 
when the number of children born per woman is less than  $2$, the total population decreases to zero. 
\end{abstract}

\maketitle

\section{Introduction}

In human history, technological advances are the main factors for human population increase, such as tool-making revolution, agricultural revolution and industrial revolution. Technological advances provide human with more food supply and medicine. 
More food supply is the main driver of population increase. Medicine prolongs human life span. 
Diseases such as plagues could cause human population to temporarily decrease. But since 1700, human population has been monotonically increasing due to technological advances. Since 1960s, due to the introduction of high yield grains, agricultural 
machineries, fertilizers, chemical pesticides, and better irrigation systems, human population has been increasing by 1 billion every 12 years, from 3 billion to 8 billion by 2023. Thus enrichment of food has increased human population dramatically. It seems that both the Cornucopian and the Malthusian views were realized \cite{Kha77} \cite{Mal26}. Human indeed dramatically advanced technology to provide abundant food supply to meet the demand of population growth according to Cornucopian view. Human population also dramatically increased with the abundant food supply according to Malthusian view. The question is whether or not we are heading to a new Malthusian catastrophe, i.e. some people are going to starve. Technologies may be advanced further to support more humans. But the earth 
resource is limited, and the human population cannot increase without bound on earth.

Human 
overpopulation not only can cause huge damage to earth resource and environment, but also has serious sustainability consequence. 
If there is a global food scarcity, huge famine can cause major population loss. According to World Wide Fund for Nature \cite{WWF06}, the current human population is already exceeding its earth carrying capacity. On the other hand, estimating earth's carrying capacity for human is more difficult than for other animals due to the fact that human choices may play an important role \cite{Coh95}. In the long run, human population cannot continue to grow; there are clear human resource limits of food, energy and territory (individual human space) as discussed by von Hoerner \cite{Hoe75}. The key moment is when human population reaches its maximum. The crucial question is: How will the human population change afterward? Will human population more or less stay at a stagnation population or decrease substantially? If human population 
decreases, is the decrease due to birth control, normal death or abnormal death? Birth control and normal death are hopeful for reducing human population from the example of China. Abnormal death corresponds to various kinds of disasters such as diseases, wars etc.. Von Hoerner also proposed the possibility of moving humans out of earth, i.e. stellar expansion \cite{Hoe75}. But wars and diseases are more probable. 

There have been a lot of effort in fitting human population historical data with a function such as the nice fitting by $c(t_* - t)^{-\al}$ for positive parameters $c$, $t_*$ and $\al$ \cite{KM16}. There were also studies on human population dynamics from 
logistic point of view \cite{MMA96} and ecological perspectives \cite{HL19}. 

Here we are focusing on the temporal evolution of the population distribution in age, and introduce the population-age-time model.

\section{The population-age-time (PAT) model}

Let $p(t,a)$ be the population of age $a$ (in year) at time $t$ (in year). One can think $p(t,a)$ as the spectrum of population in 
age, $a = 0, 1, 2, \cdots , A$; $t=0,1,2 \cdots $. We introduce the following population-age-time (PAT) model,
\begin{eqnarray}
p(t+1,a) &=& p(t,a-1) - \om (t, a-1) p(t,a-1), \quad a=1,2,\cdots , A; \label{PAT1} \\
p(t+1,0) &=& \sum_{b=B_1}^{B_2} \al (t,b) p(t,b) ;  \label{PAT2} 
\end{eqnarray}
where $\om (t, a-1)$ is the death rate of the population $p(t,a-1)$, $\al (t,b)$ is the birth rate of the population $p(t,b)$, 
$p(t,A)$ is set to zero, and $A$, $B_1$ and $B_2$ are e.g. 
\begin{equation}
A=120, \  B_1 = 18, \  B_2 = 40 . \label{sc}
\end{equation}
Here the effects of immigration/emigration and migration are not included. 
The total birth rate 
\begin{equation}
\al = \sum_{b=B_1}^{B_2} \al (t, b) , \label{tbr}
\end{equation}
roughly represents half of the number of children born per woman (under the assumption that male and female populations are roughly the same). For Europe, $\al$ is about $0.7$; for Africa, $\al$ is
$2$ to $2.5$; and for China, $\al$ is about $0.5$. Factors that can affect birth rate include food supply, medical technology,
social norm, and affordability (expense). Factors that can affect death rate include age, disease (medical technology), 
food supply, war, accident, disaster. 

The evolution of the total population
\begin{equation}
P(t) = \sum_{a=0}^A p(t,a) , \label{tp}
\end{equation}
satisfies
\[
P(t+1) = P(t) +\sum_{b=B_1}^{B_2} \al (t,b) p(t,b) - \sum_{a=1}^A \om (t, a-1) p(t,a-1),
\]
where the second term is the total birth and the last term is the total death in the year $t$. 

The steady state population distribution $p(t,a) = p(a)$, ($a=0, 1, 2, \cdots, A$) satisfies
\begin{eqnarray}
p(a) &=& p(a-1) - \om (a-1) p(a-1), \quad a=1,2,\cdots , A; \label{sp1} \\
p(0) &=& \sum_{b=B_1}^{B_2} \al (b) p(b) ;  \label{sp2} 
\end{eqnarray}
thus
\begin{eqnarray*}
p(a) &=& [1 - \om (a-1)] \cdots [1 - \om (0)] p(0), \quad a=1,2,\cdots , A;  \\
p(0) &=& \sum_{b=B_1}^{B_2} \al (b)[1 - \om (b-1)] \cdots [1 - \om (0)] p(0)  ;   \\
\end{eqnarray*}
and we arrive at the following constraint on $\al (b)$ and $\om (b)$:
\[
1= \sum_{b=B_1}^{B_2} \al (b)[1 - \om (b-1)] \cdots [1 - \om (0)] .
\]
We are interested in the dynamics of the population distribution and the total population, and its dependence upon birth/death rate
distribution and initial conditions. 

\section{Robustness of the population distribution dynamics}

\begin{figure}[ht] 
\centering
\subfigure[linear birth rate]{\includegraphics[width=2.3in,height=2.3in]{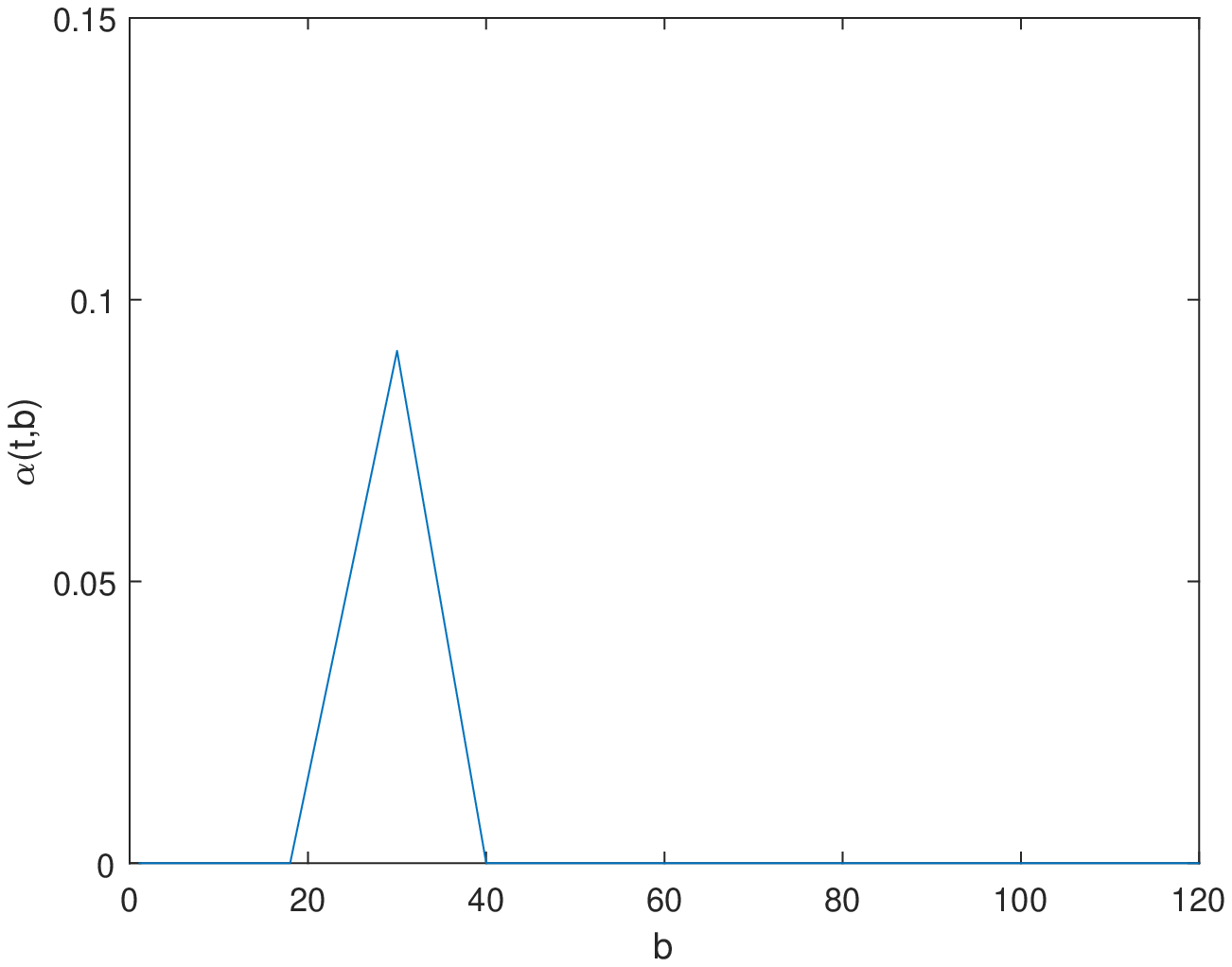}}
\subfigure[linear death rate]{\includegraphics[width=2.3in,height=2.3in]{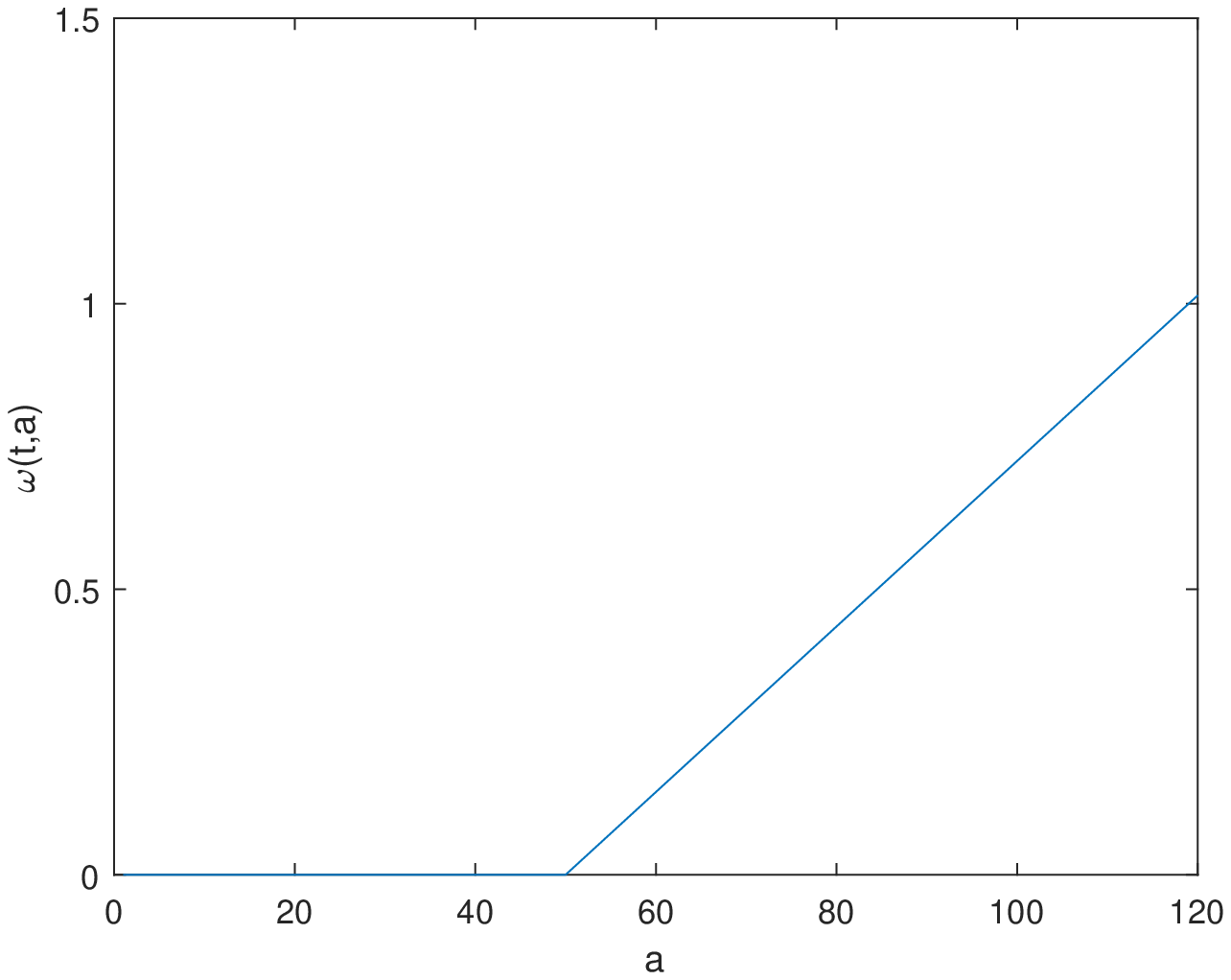}}
\caption{(a). The graphical illustration of the linear birth rate (\ref{simbr}). (b). The graphical illustration of the linear death rate (\ref{simdr}).}
\label{flbdr}
\end{figure}

With the choice of parameters (\ref{sc}), we start with the following piecewise linear birth and death rates:
\begin{equation}
\al (t,b) = \left \{ \begin{array}{lr} \frac{\e}{12}(b-18),  & 18 \leq b \leq 30, \cr 
\frac{\e}{10}(40-b), & 30 < b \leq 40 ,
\end{array} \right .
\label{simbr}
\end{equation}
\begin{equation}
\om (t,a) = \left \{ \begin{array}{lr} 0,  & 0 \leq a \leq n, \cr 
\frac{1}{119-n}(a-n), & n < b \leq 119 .
\end{array} \right .
\label{simdr}
\end{equation}
See Figure \ref{flbdr} for a graphical illustration. These are the simplest birth rate and death rate that we can think of. The key 
point is that we are going to show that the qualitative nature of the population distribution dynamics is robust with respect to 
different forms of birth rate and death rate distributions in age. 

\begin{figure}[ht] 
\centering
\subfigure{\includegraphics[width=2.3in,height=2.3in]{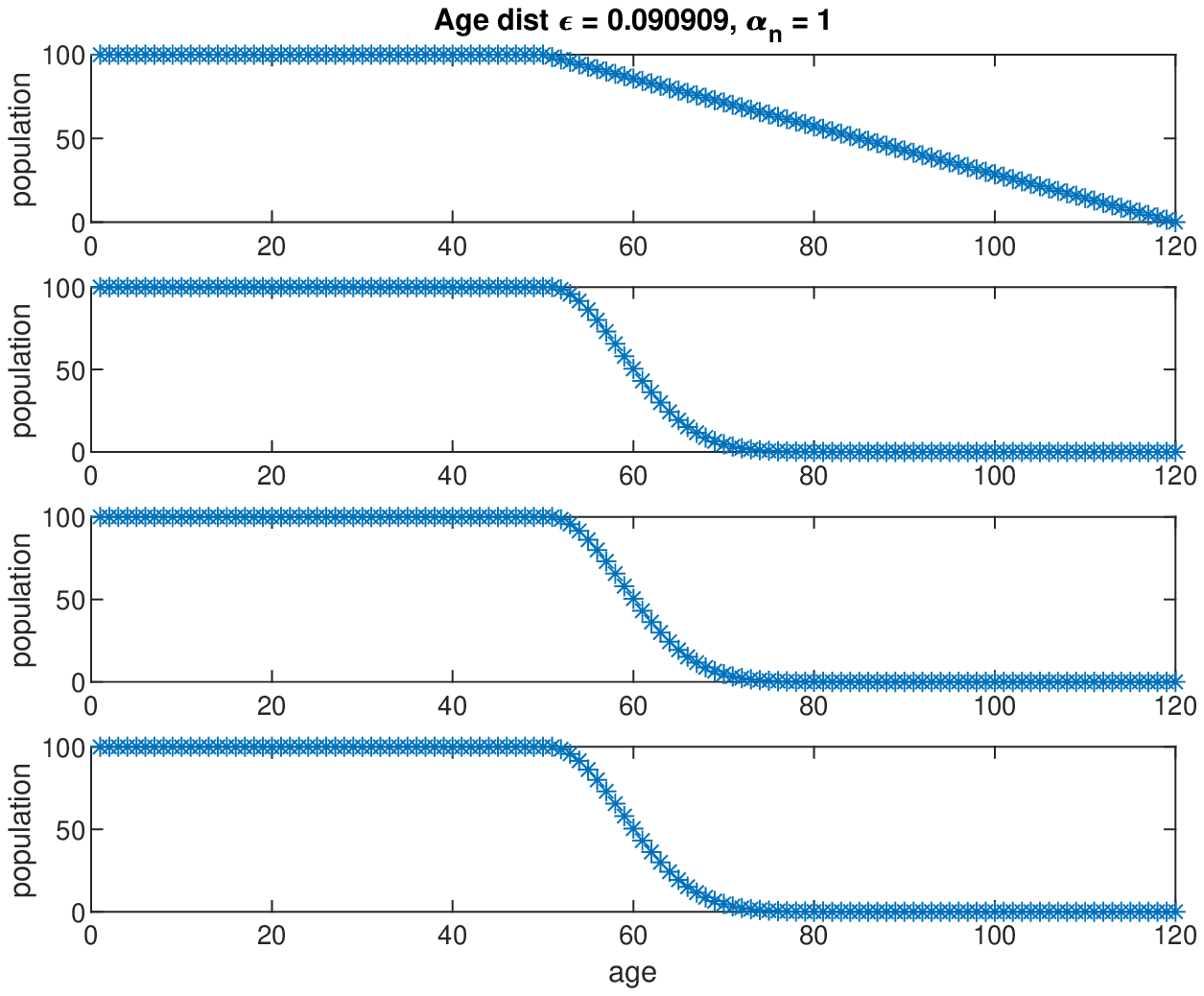}}
\subfigure{\includegraphics[width=2.3in,height=2.3in]{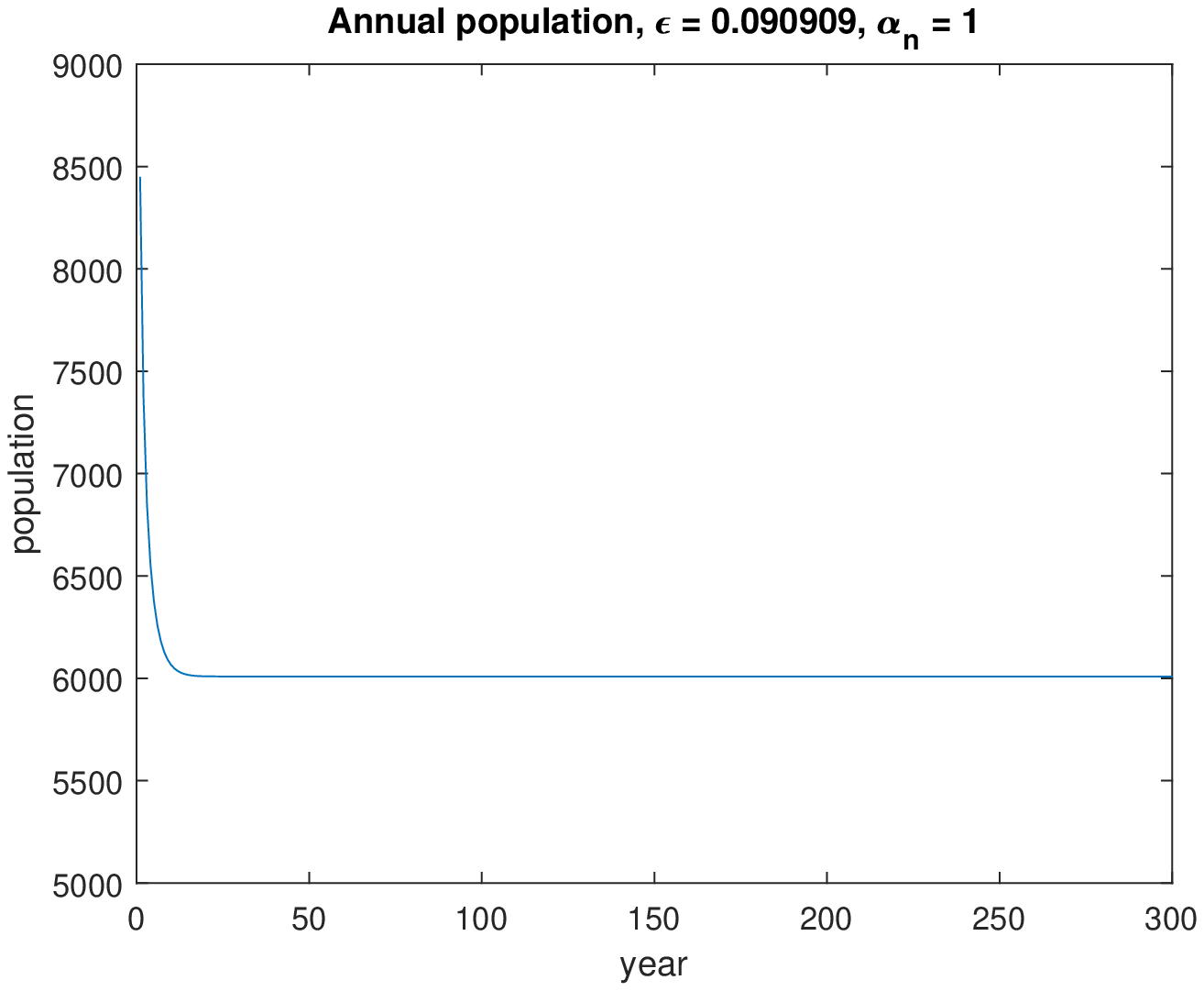}}
\subfigure{\includegraphics[width=2.3in,height=2.3in]{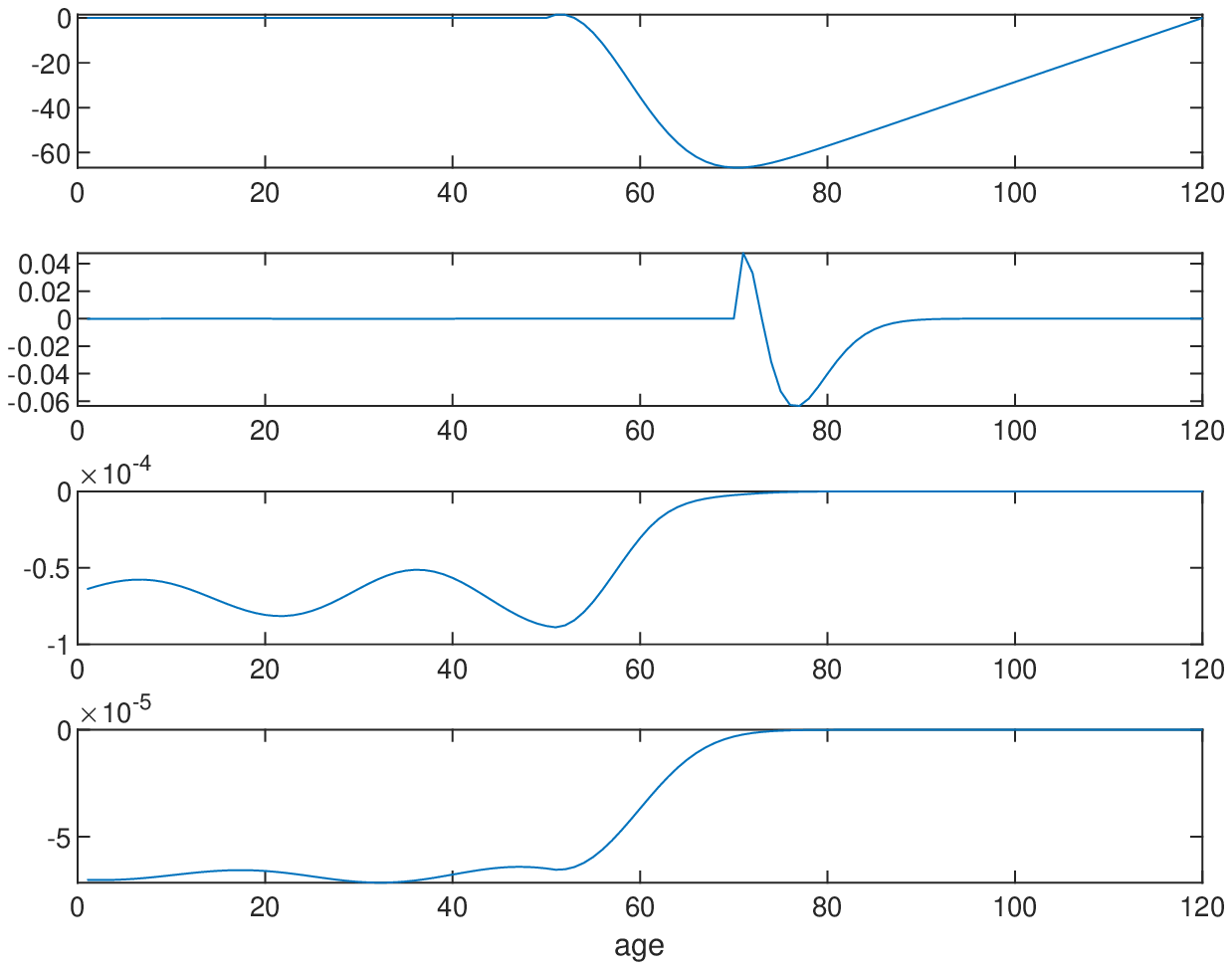}}
\caption{(a). From the top to the bottom is the temporal evolution of the population distribution: $p(0,a)$; $p(20,a)$; $p(100,a)$; $p(200,a)$. The total birth rate $\al = \al_n = 1$ for $n=50$. (b). The temporal evolution of the total population $P(t)$. (c). From the top to the bottom is the temporal evolution of the population distribution difference: $p(20,a)-p(0,a)$; 
$p(40,a)-p(20,a)$; $p(120,a)-p(100,a)$; $p(220,a)-p(200,a)$.}
\label{A1}
\end{figure}

\begin{figure}[ht] 
\centering
\subfigure{\includegraphics[width=2.3in,height=2.3in]{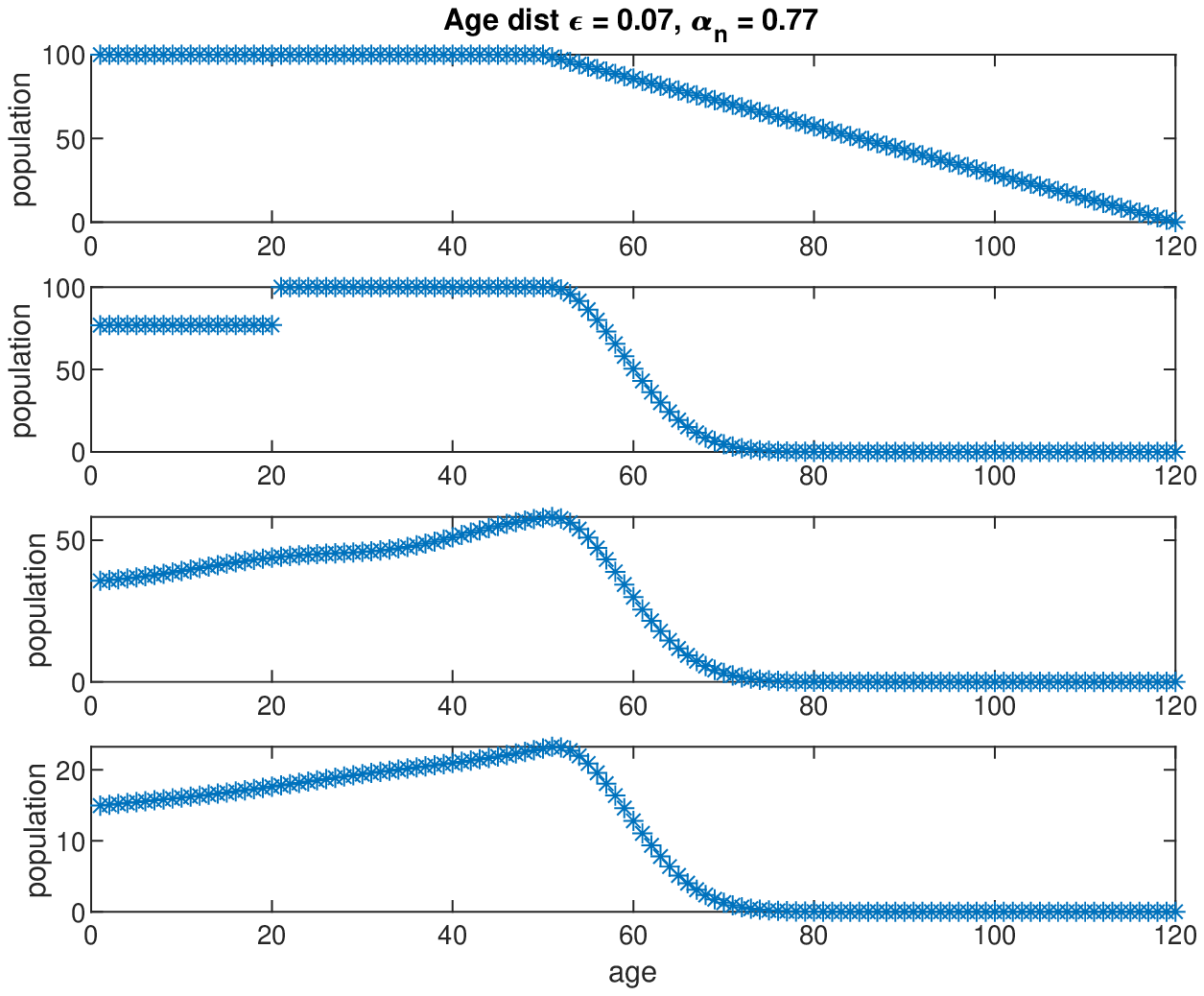}}
\subfigure{\includegraphics[width=2.3in,height=2.3in]{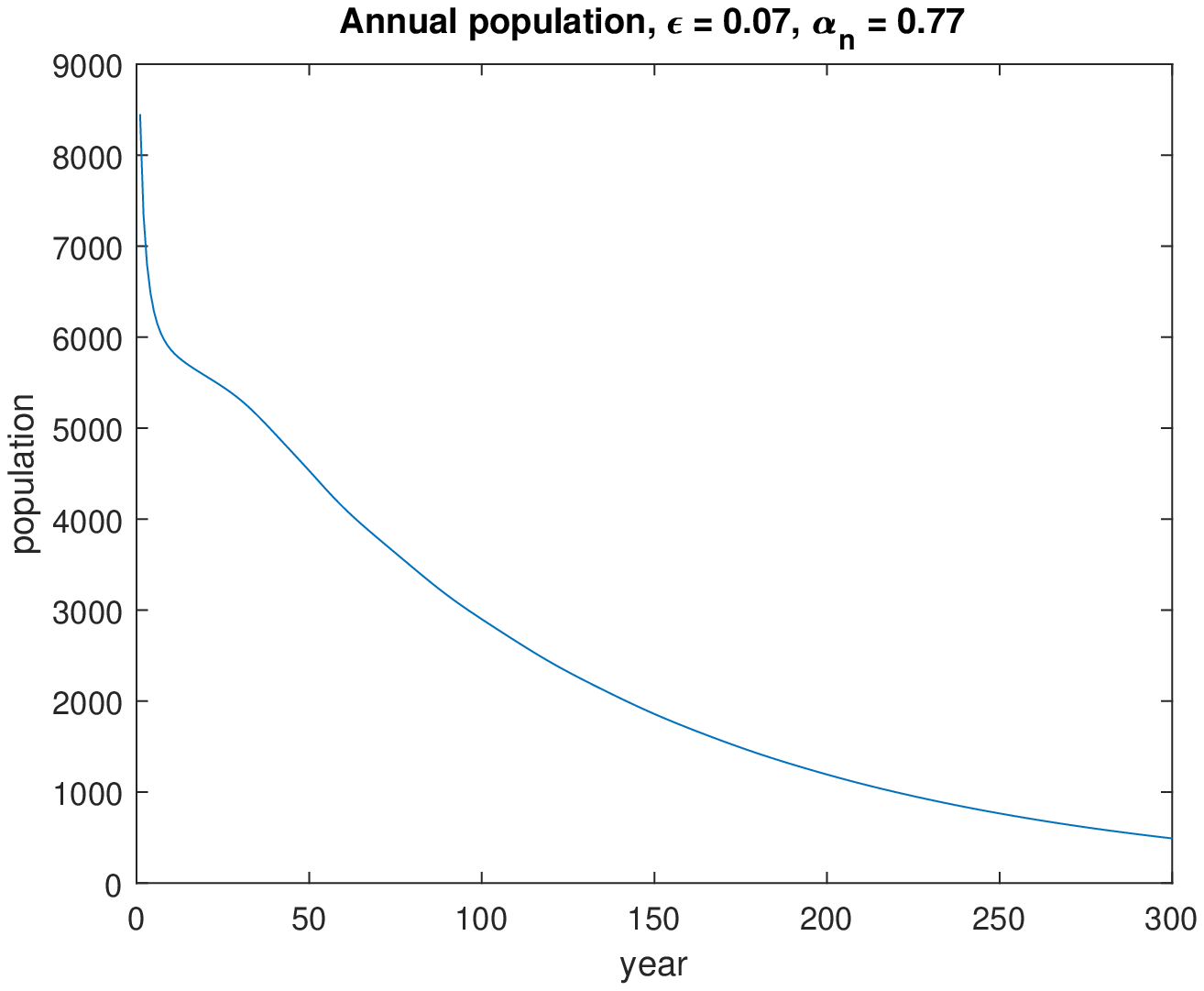}}
\subfigure{\includegraphics[width=2.3in,height=2.3in]{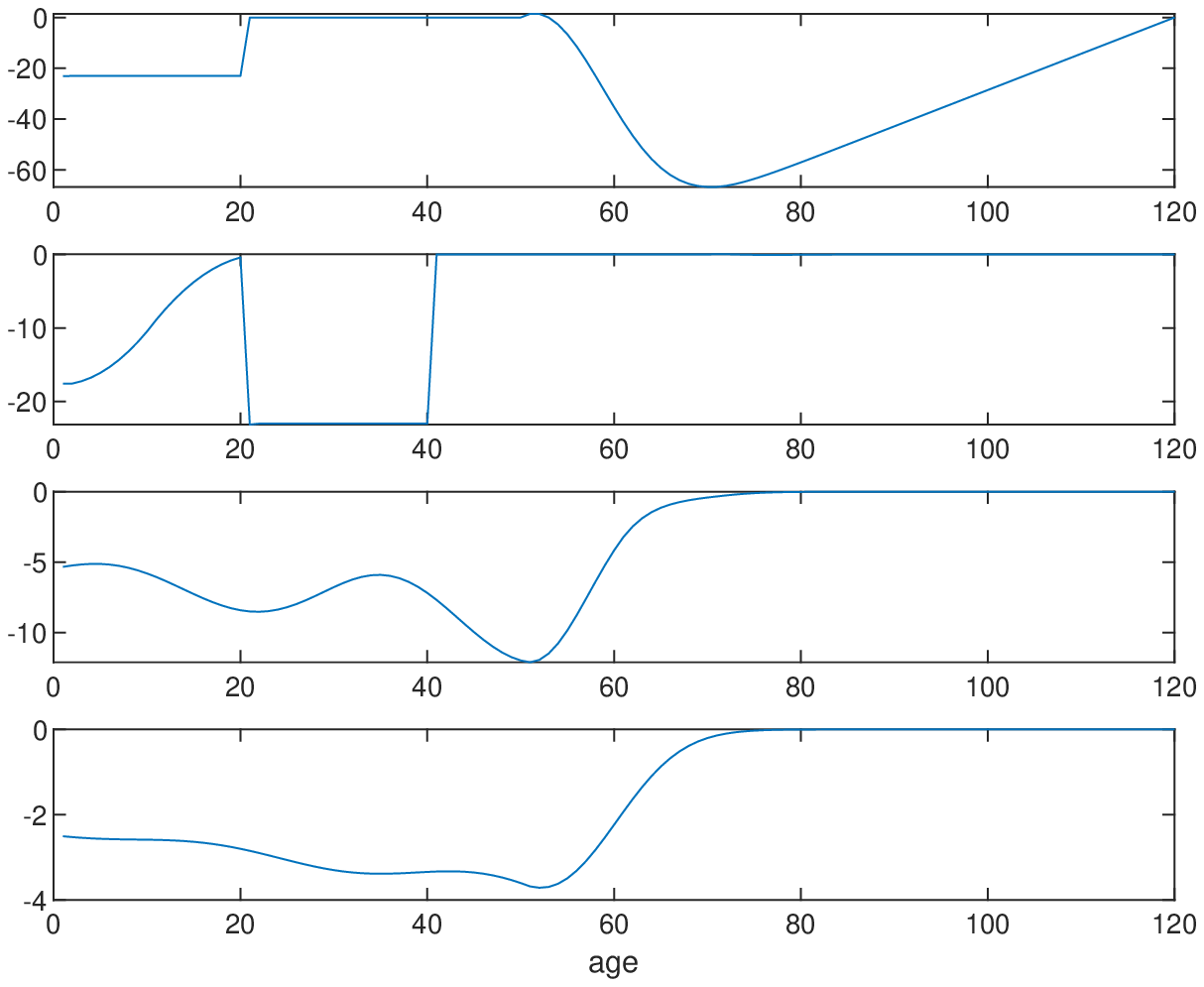}}
\caption{(a). From the top to the bottom is the temporal evolution of the population distribution: $p(0,a)$; $p(20,a)$; $p(100,a)$; $p(200,a)$. The total birth rate $\al = \al_n = 0.77$ for $n=50$. (b). The temporal evolution of the total population $P(t)$. (c). From the top to the bottom is the temporal evolution of the population distribution difference: $p(20,a)-p(0,a)$; 
$p(40,a)-p(20,a)$; $p(120,a)-p(100,a)$; $p(220,a)-p(200,a)$.}
\label{A77}
\end{figure}

\begin{figure}[ht] 
\centering
\subfigure{\includegraphics[width=2.3in,height=2.3in]{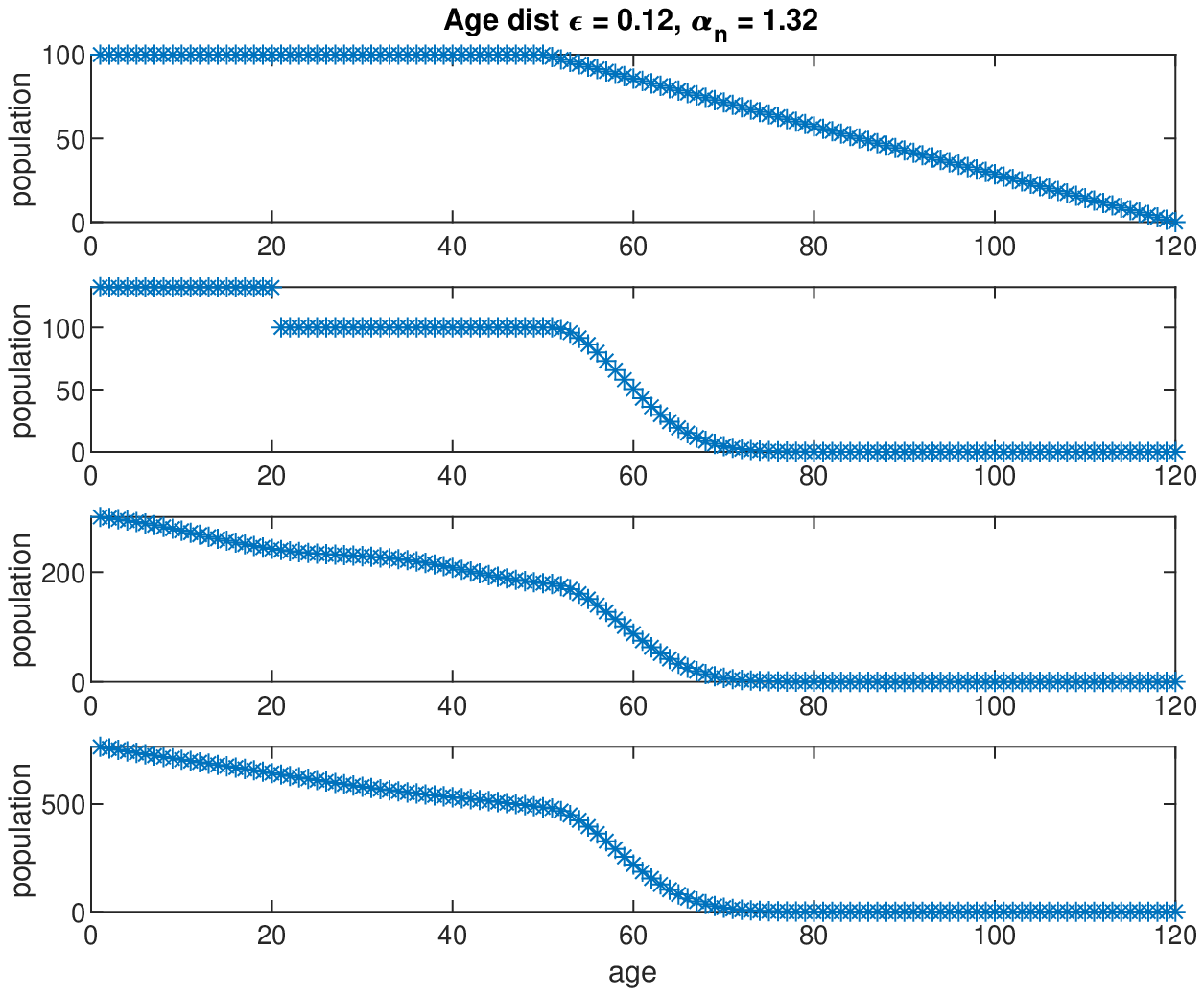}}
\subfigure{\includegraphics[width=2.3in,height=2.3in]{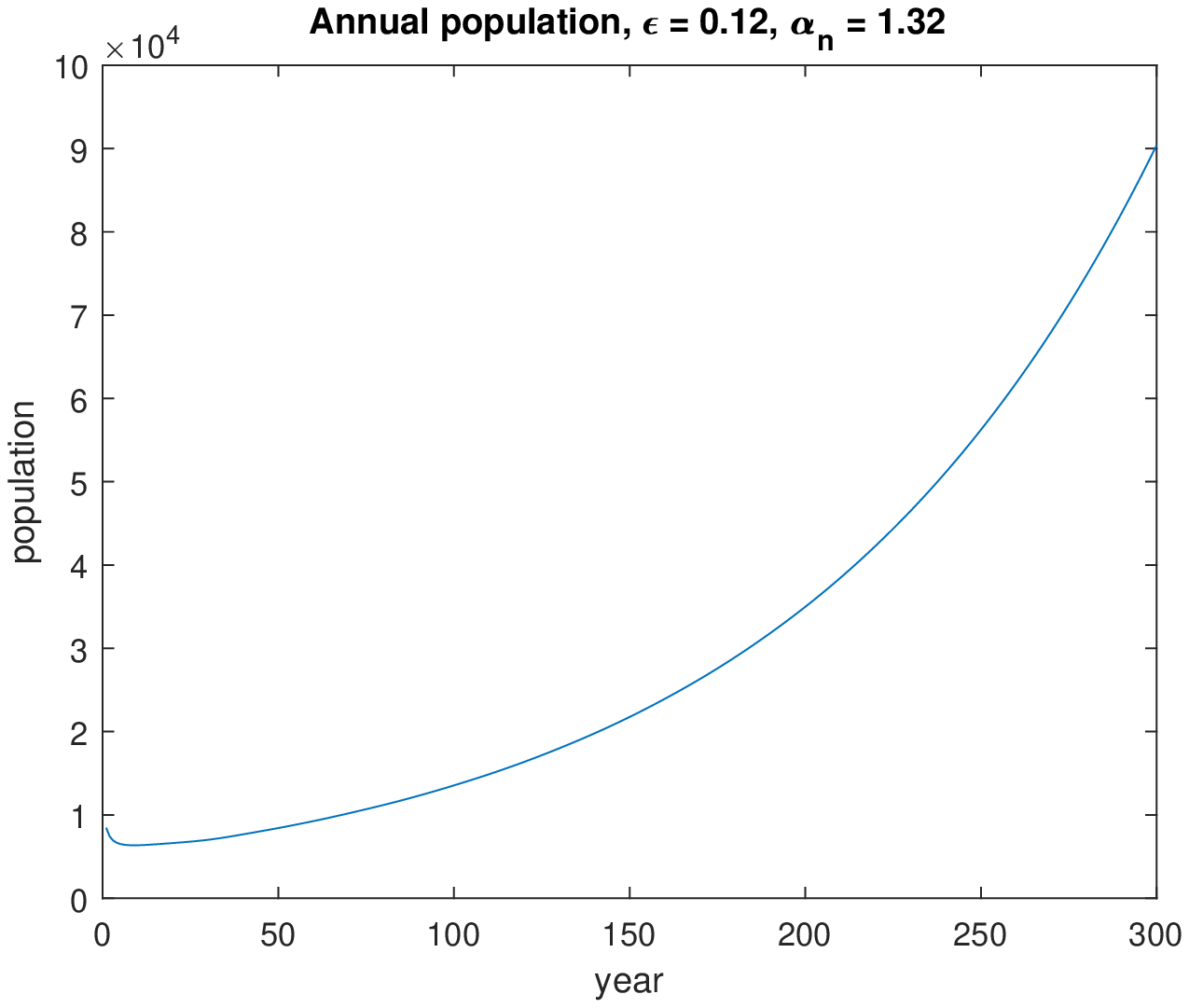}}
\subfigure{\includegraphics[width=2.3in,height=2.3in]{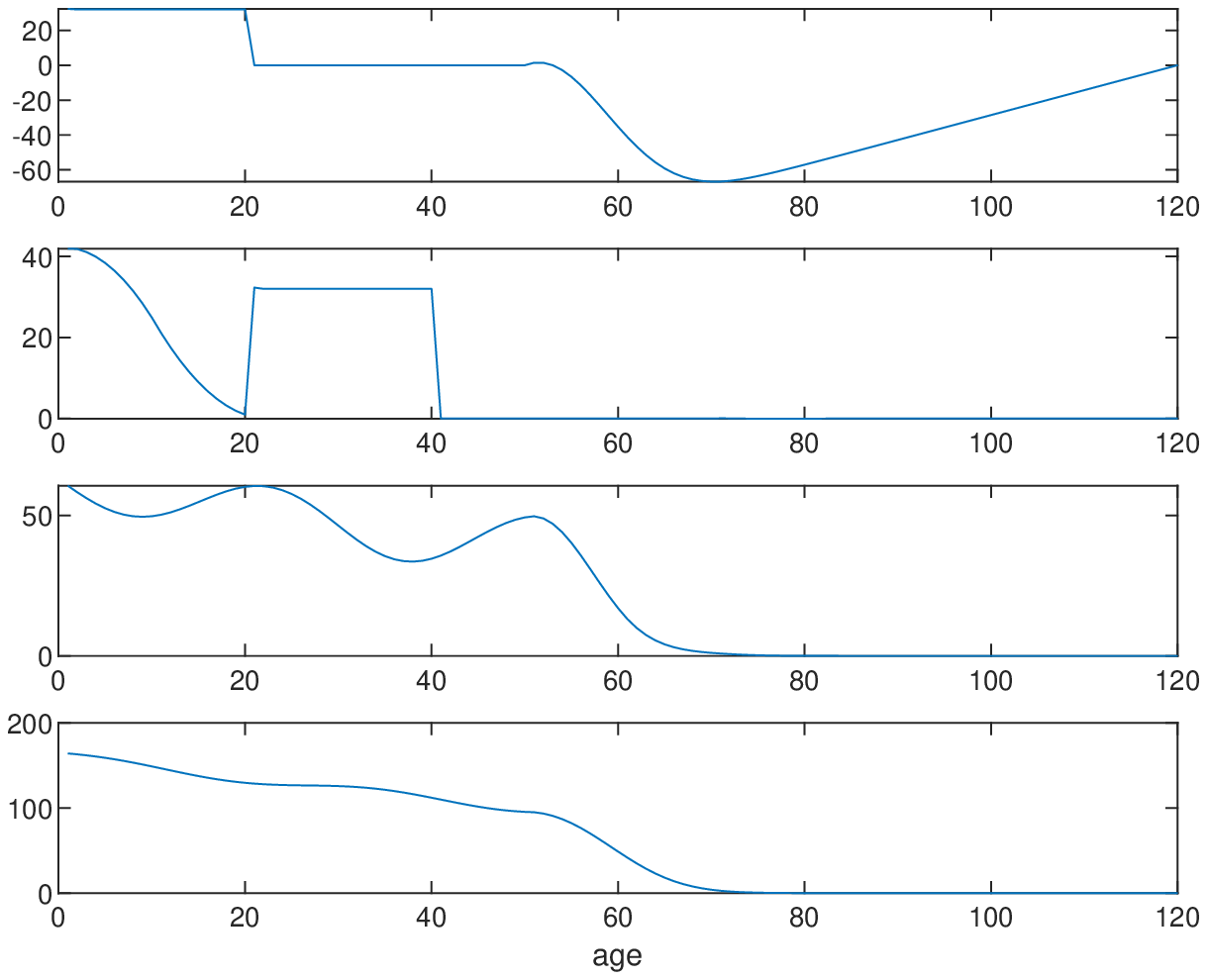}}
\caption{(a). From the top to the bottom is the temporal evolution of the population distribution: $p(0,a)$; $p(20,a)$; $p(100,a)$; $p(200,a)$. The total birth rate $\al = \al_n = 1.32$ for $n=50$. (b). The temporal evolution of the total population $P(t)$. (c). From the top to the bottom is the temporal evolution of the population distribution difference: $p(20,a)-p(0,a)$; 
$p(40,a)-p(20,a)$; $p(120,a)-p(100,a)$; $p(220,a)-p(200,a)$.}
\label{A132}
\end{figure}

The total birth rate is given by
\[
\al = \sum_{b=18}^{40} \al (t,b) = \sum_{b=18}^{30} \frac{\e}{12}(b-18) +  \sum_{b=31}^{40}\frac{\e}{10}(40-b) .
\]
Choosing $n=50$, the temporal dynamics of the population is shown in Figures \ref{A1} - \ref{A132}. The most interesting 
feature is the development of a sharp transition region right after the death rate starts to be nonzero. That is, old age population sharply declines with aging due to death. The sharp transition region makes the population distribution bearing a kink shape. 
We will show later that the kink shape is universal with respect to different forms of birth and death rates. 
When $\al = 1$ (roughly two children born per woman), the population distribution $p(t,a)$ approaches an asymptotically steady state (see Figure \ref{A1}).  The steady state bears the typical kink shape with the age region before the sharp transition 
being constant. When $\al < 1$, the total population decreases to zero (see Figure \ref{A77}). The left portion of the kink
shape population distribution is an increasing function in age. When $\al > 1$, the total population increases without bound 
(see Figure \ref{A132}). The left portion of the kink shape population distribution is a decreasing function in age. 

\begin{figure}[ht] 
\centering
\subfigure{\includegraphics[width=2.3in,height=2.3in]{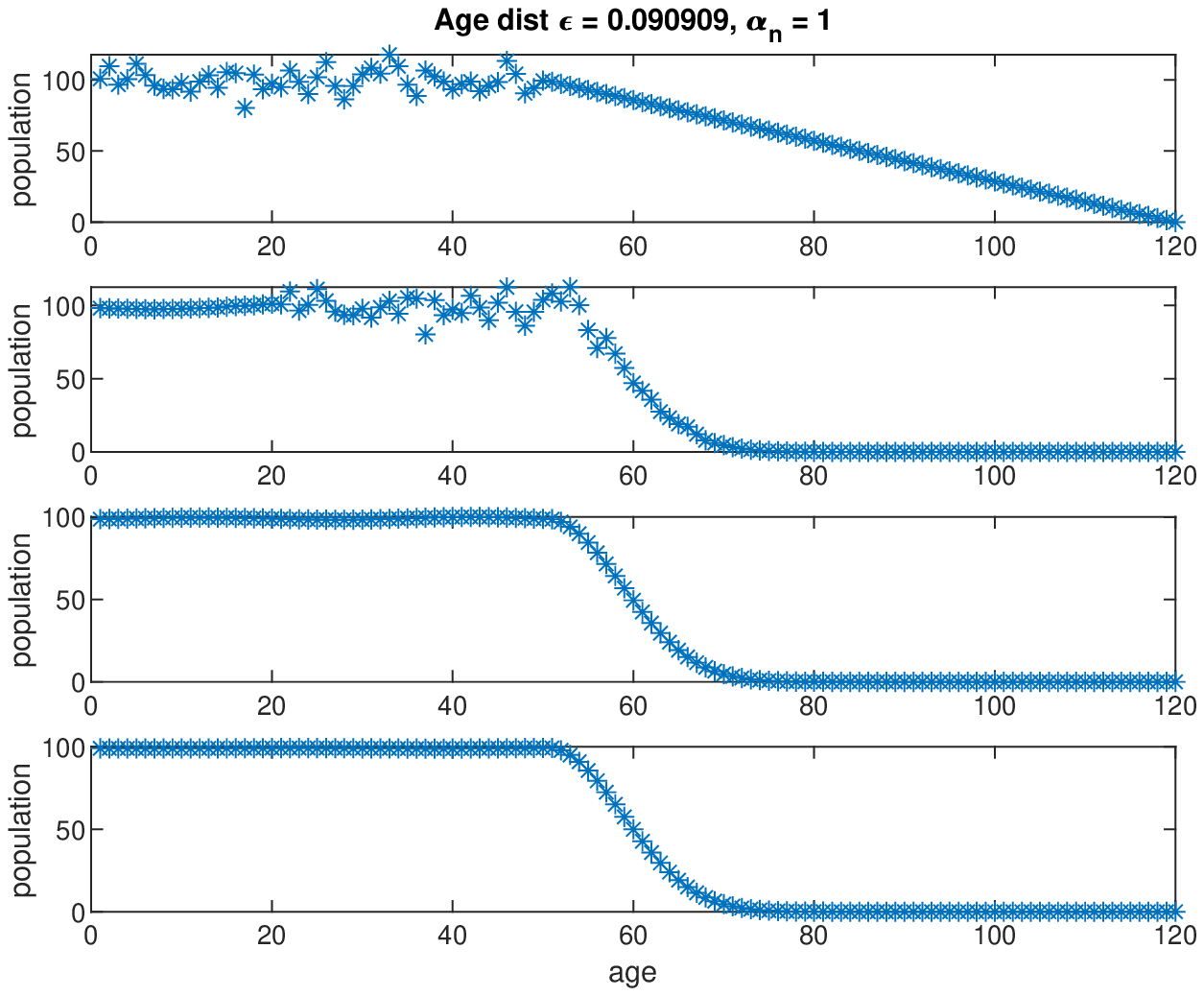}}
\subfigure{\includegraphics[width=2.3in,height=2.3in]{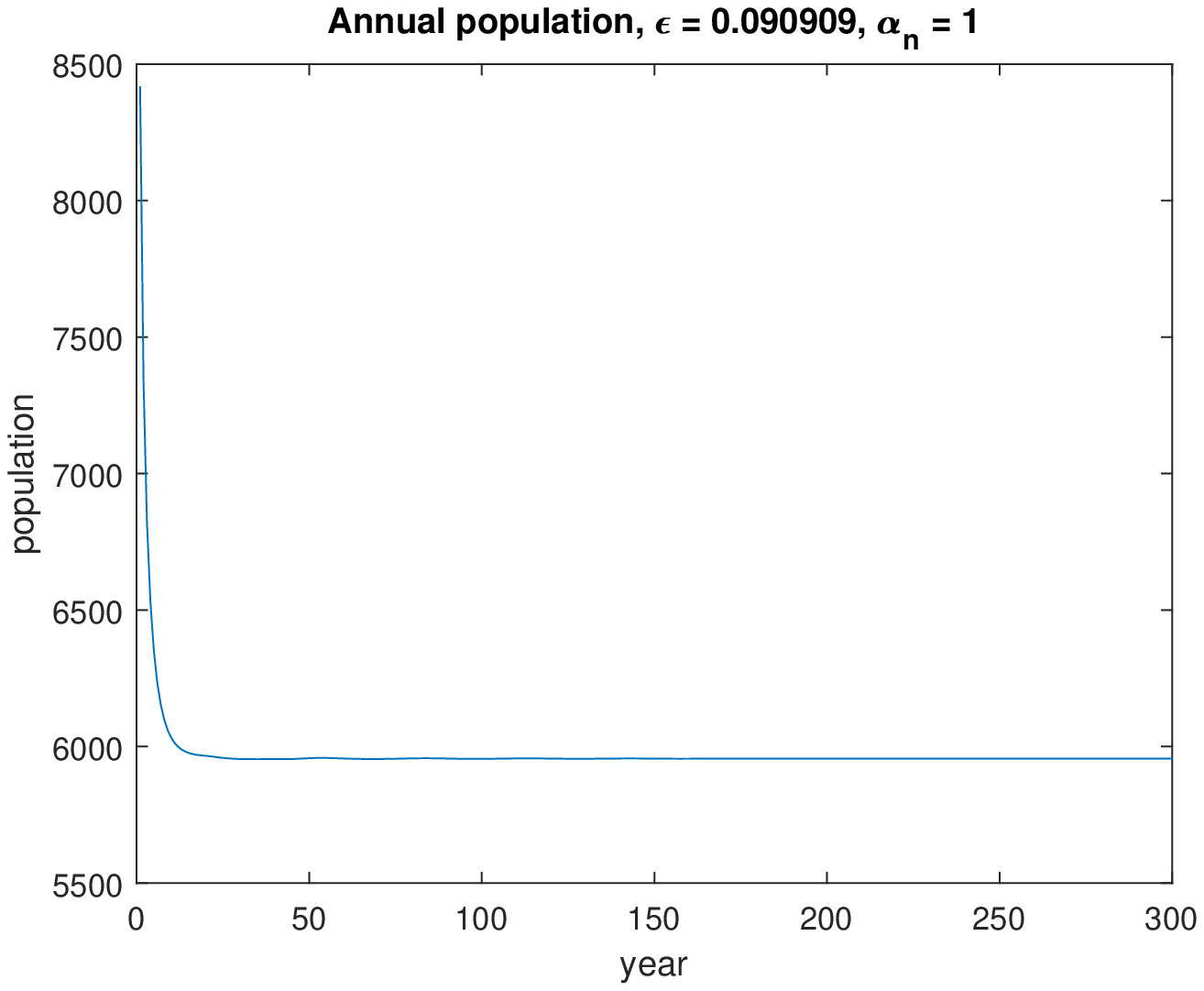}}
\subfigure{\includegraphics[width=2.3in,height=2.3in]{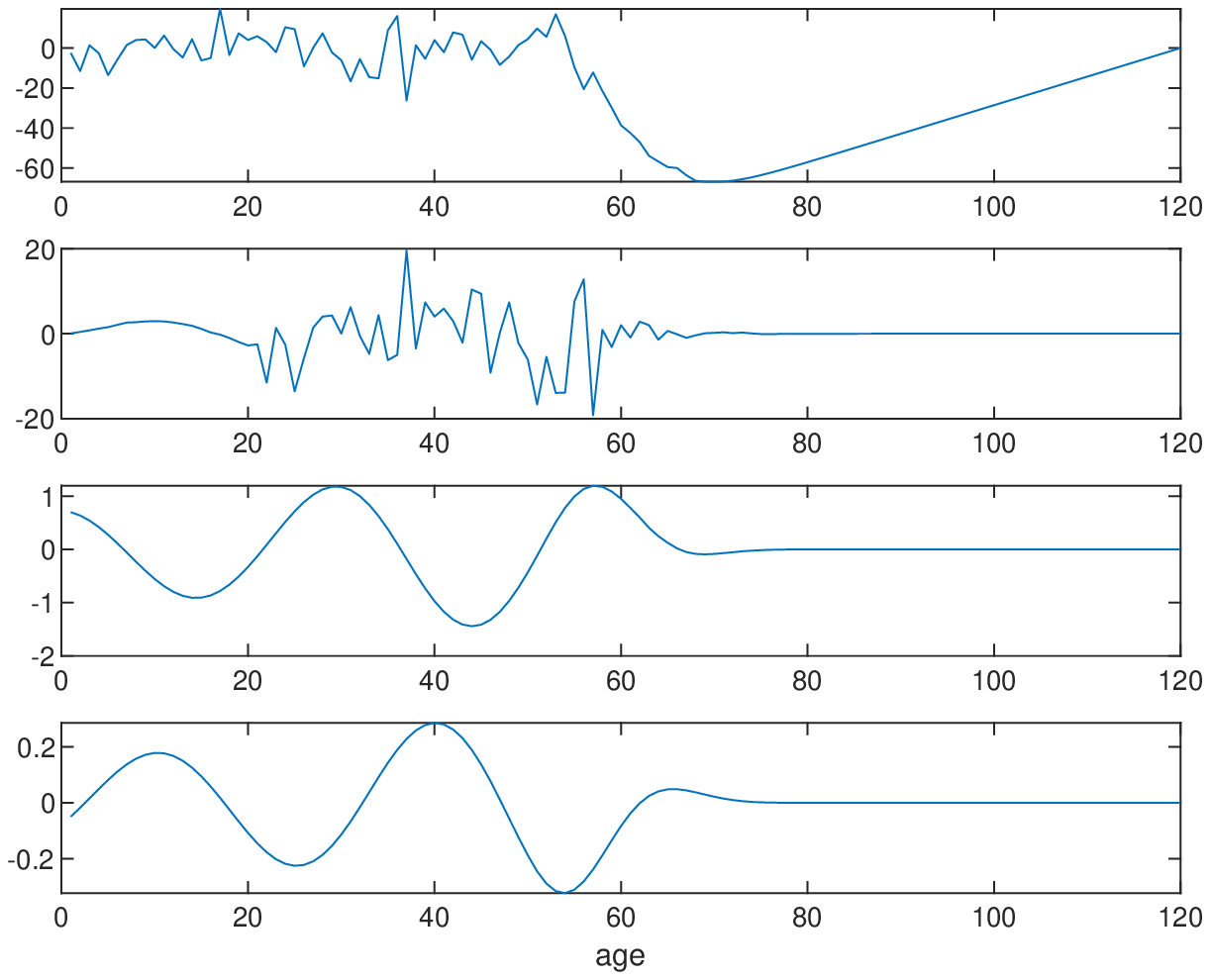}}
\caption{(a). With zero mean random perturbation on the initial condition of Figure \ref{A1}, from the top to the bottom is the temporal evolution of the population distribution: $p(0,a)$; $p(20,a)$; $p(100,a)$; $p(200,a)$. The total birth rate $\al = \al_n = 1$ for $n=50$. (b). The temporal evolution of the total population $P(t)$ with zero mean random perturbation on the initial condition of Figure \ref{A1}.  (c). With zero mean random perturbation on the initial condition of Figure \ref{A1}, from the top to the bottom is the temporal evolution of the population distribution difference: $p(20,a)-p(0,a)$; 
$p(40,a)-p(20,a)$; $p(120,a)-p(100,a)$; $p(220,a)-p(200,a)$.}
\label{Random}
\end{figure}

Now we are going to show the robustness of the qualitative nature of the population distribution dynamics with respect to 
initial conditions, birth rate, and death rate. When perturbing the initial condition with zero mean random 
perturbation (with Gaussian distribution), the random perturbation is quickly washed away in time; with  $\al = 1$, 
the population distribution $p(t,a)$ approaches the same asymptotically steady state as in Figure \ref{A1}, see 
Figure \ref{Random}. When $\al = 1$, the birth dynamics (\ref{PAT2}) amounts to a weighted averaging of the fertile 
population. Such an averaging washes away the random perturbation.  

\begin{figure}[ht] 
\centering
\subfigure{\includegraphics[width=2.3in,height=2.3in]{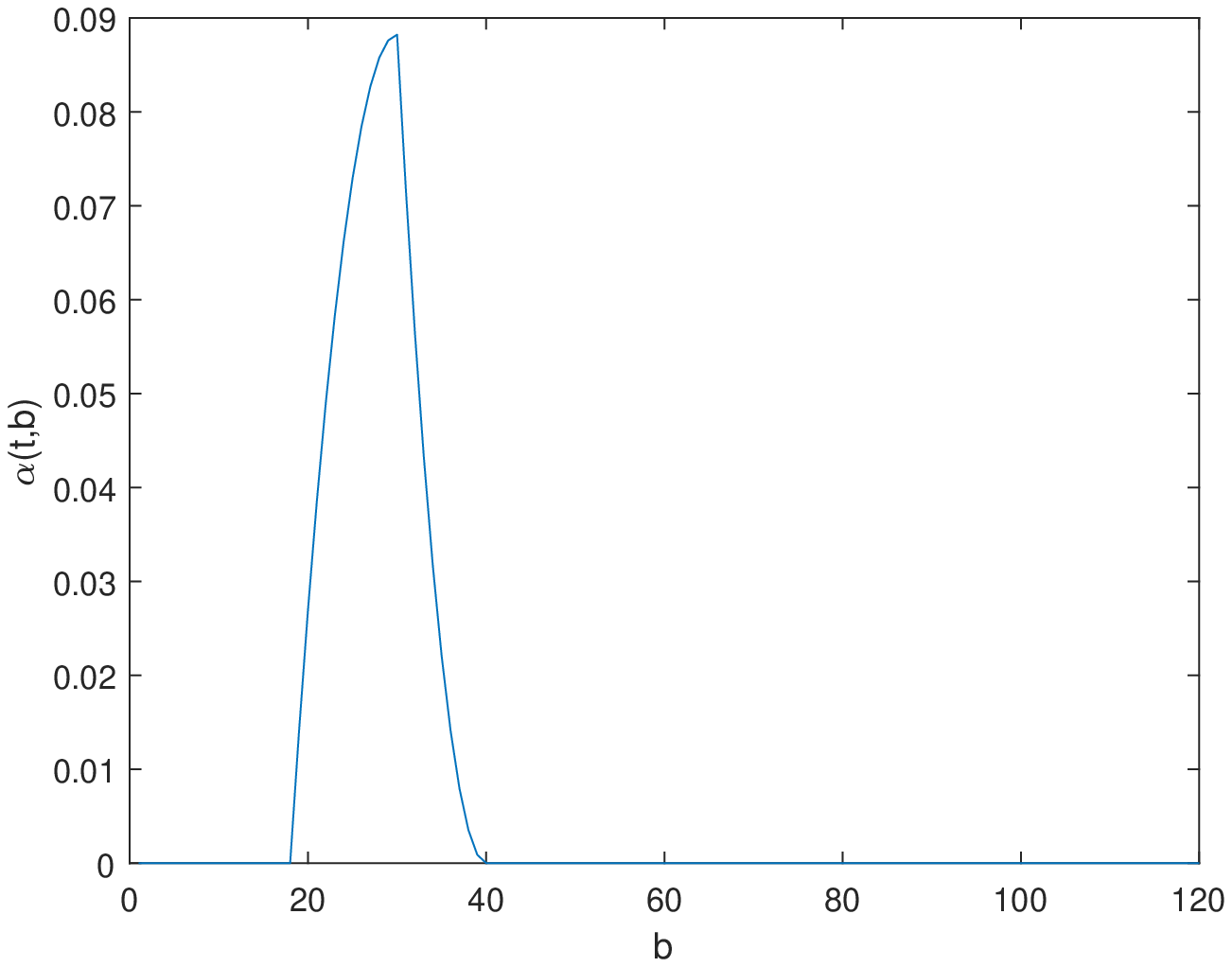}}
\subfigure{\includegraphics[width=2.3in,height=2.3in]{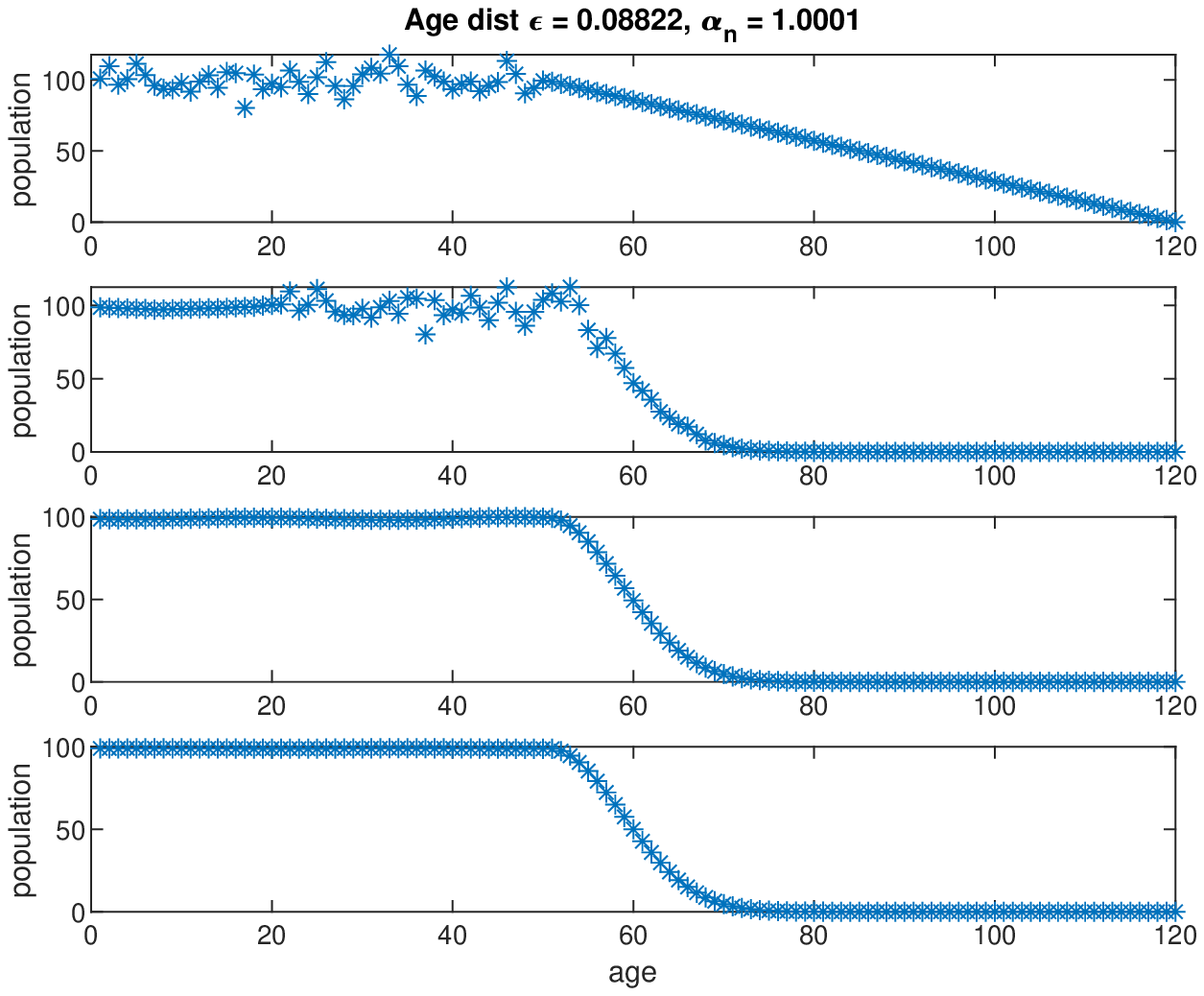}}
\subfigure{\includegraphics[width=2.3in,height=2.3in]{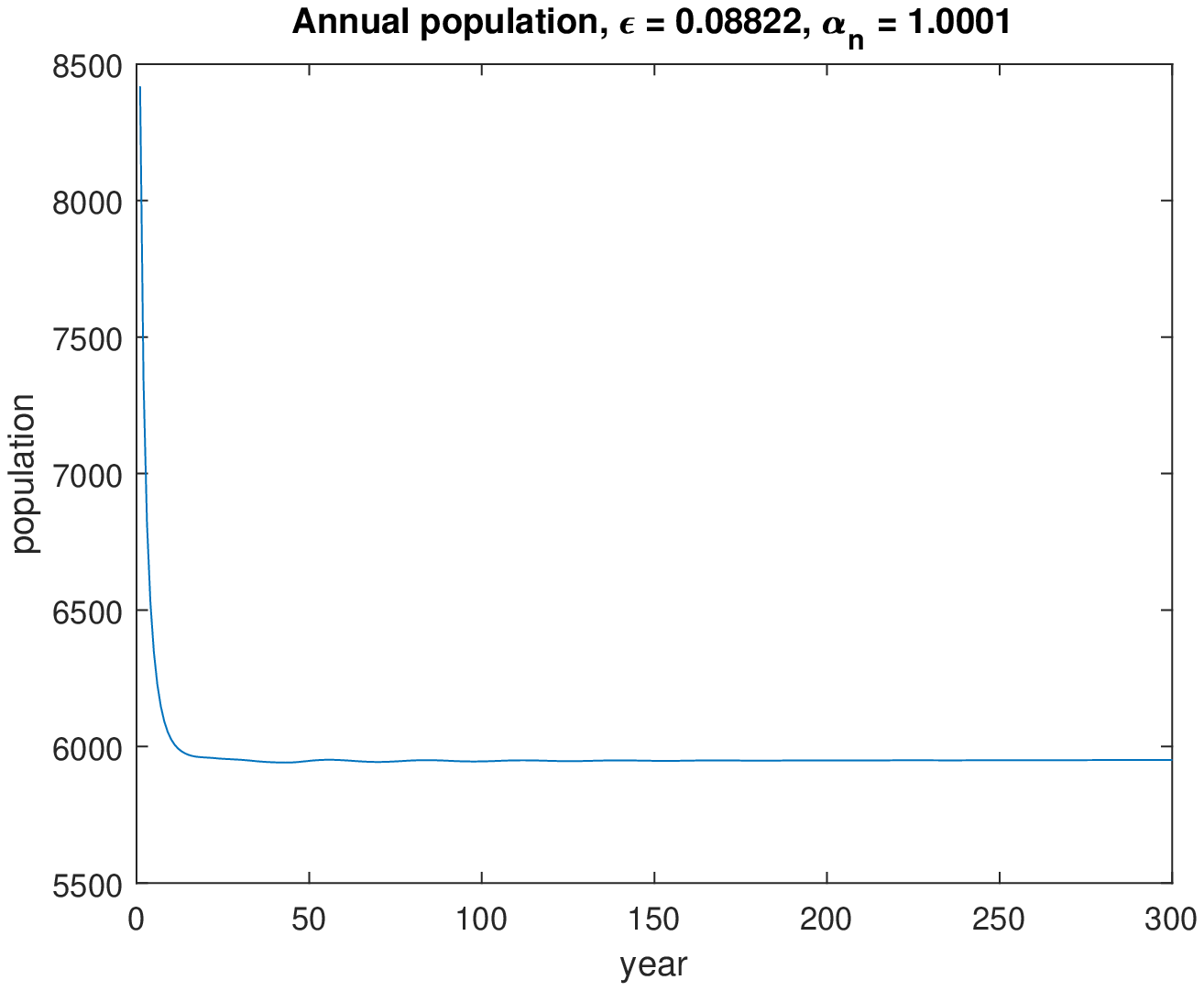}}
\subfigure{\includegraphics[width=2.3in,height=2.3in]{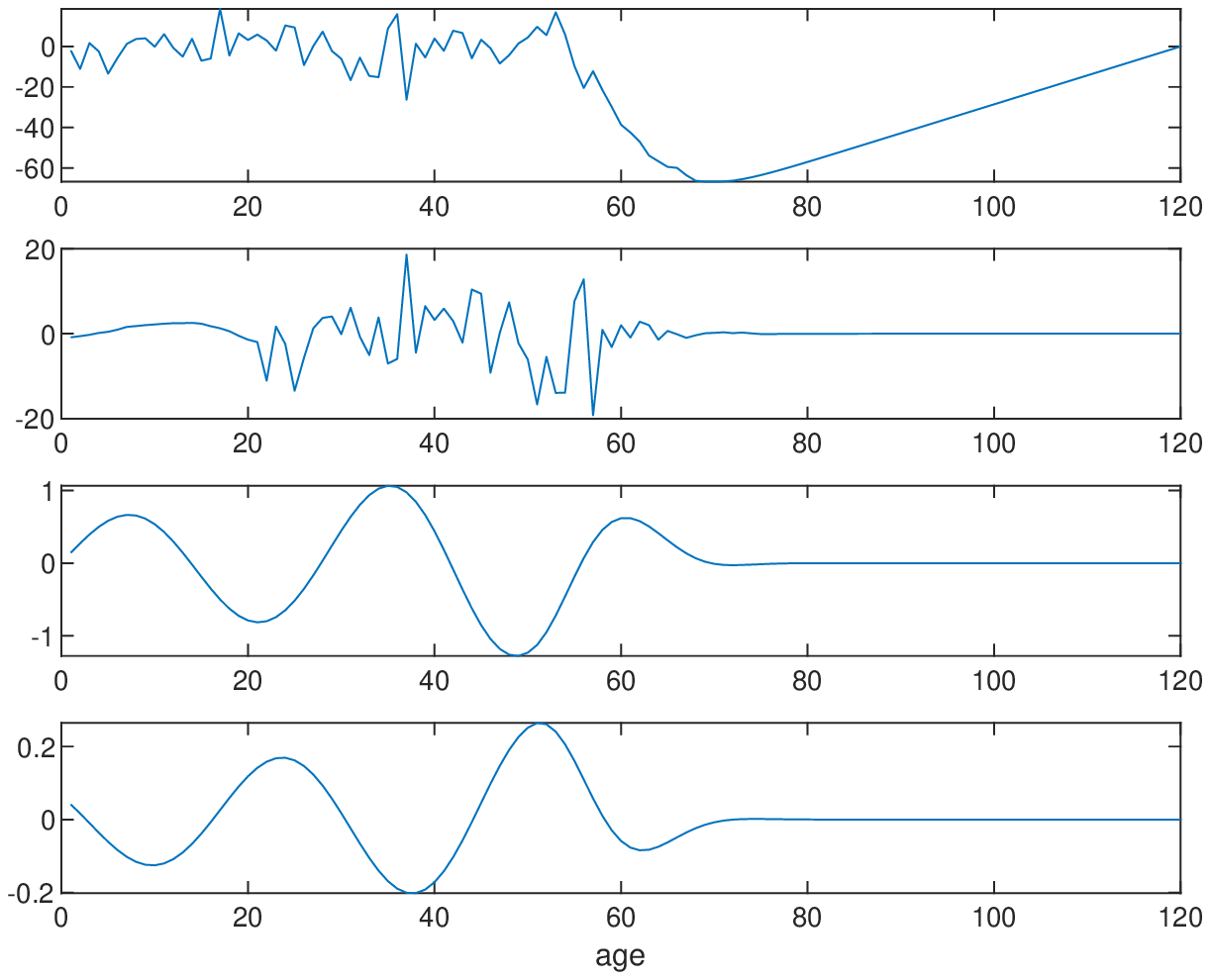}}
\caption{(a). The graphical illustration of a mixed superlinear and sublinear birth rates. (b). With the mixed birth rates in (a) and zero mean random perturbation on the initial condition of Figure \ref{A1}, from the top to the bottom is the temporal evolution of the population distribution: $p(0,a)$; $p(20,a)$; $p(100,a)$; $p(200,a)$. The total birth rate $\al = \al_n = 1.0001$ for $n=50$. (c). The temporal evolution of the total population $P(t)$. (d). Ffrom the top to the bottom is the temporal evolution of the population distribution difference: $p(20,a)-p(0,a)$; 
$p(40,a)-p(20,a)$; $p(120,a)-p(100,a)$; $p(220,a)-p(200,a)$. }
\label{mix}
\end{figure}

Replacing the linear birth or death rates with sublinear or superlinear rates does not change the nature of the dynamics (see 
Figure \ref{mix}). Since the birth dynamics amounts to a weighted averaging of the fertile population when $\al = 1$. Such an averaging does not change substantially when we replace the linear birth rates with sublinear or superlinear rates. The death rate always creates a sharp transition region in the population distribution.

\begin{figure}[ht] 
\centering
\subfigure{\includegraphics[width=2.3in,height=2.3in]{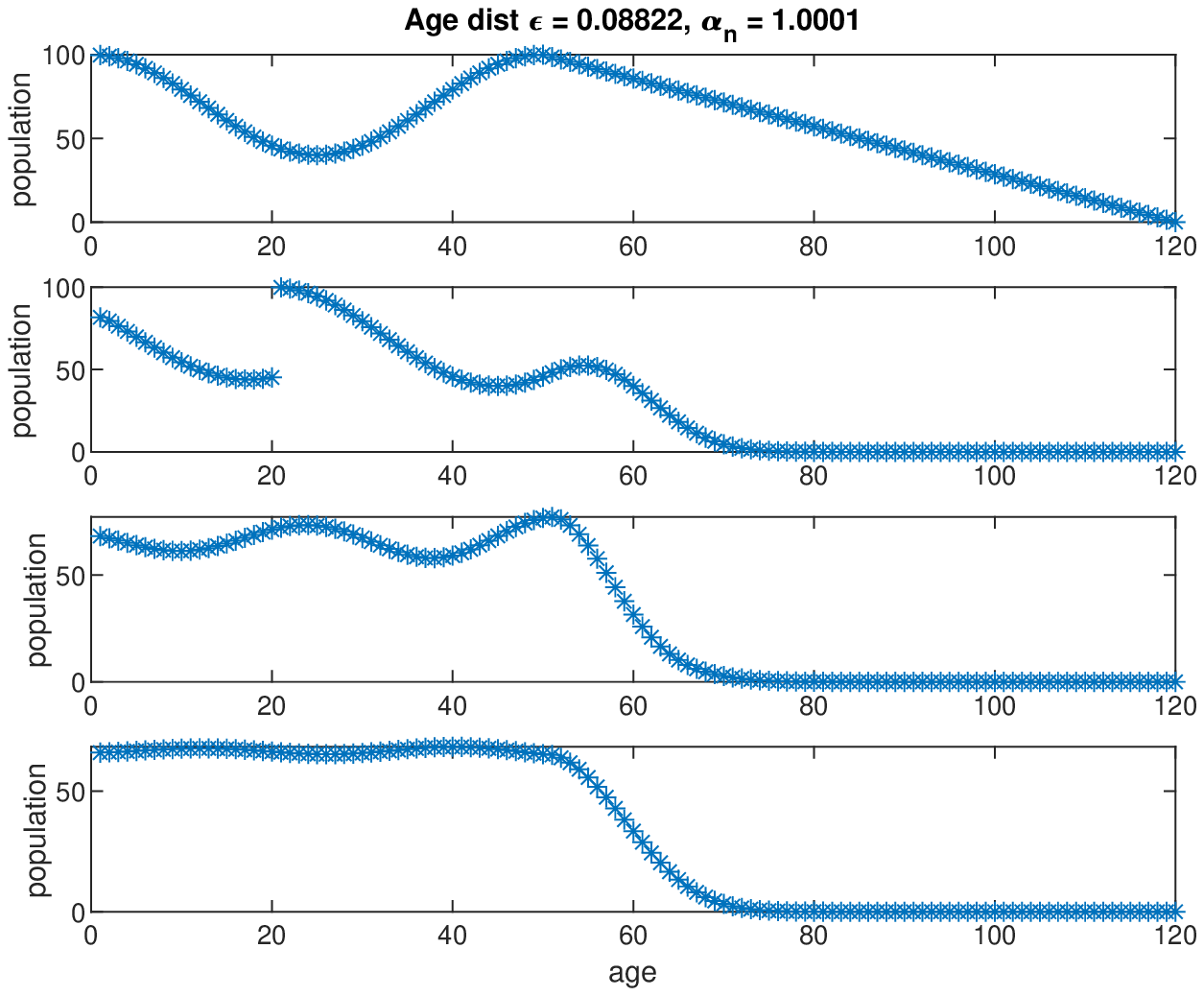}}
\subfigure{\includegraphics[width=2.3in,height=2.3in]{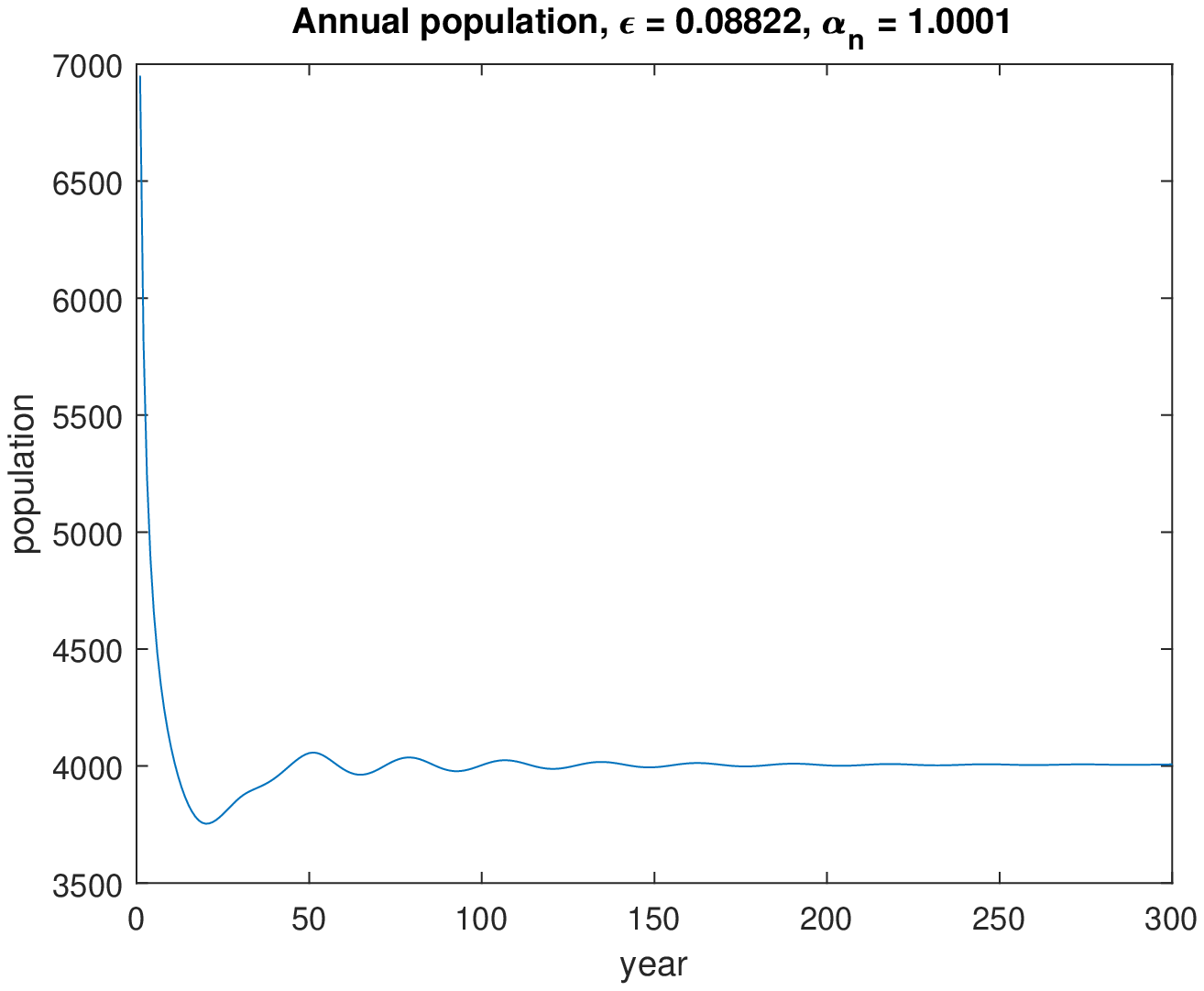}}
\subfigure{\includegraphics[width=2.3in,height=2.3in]{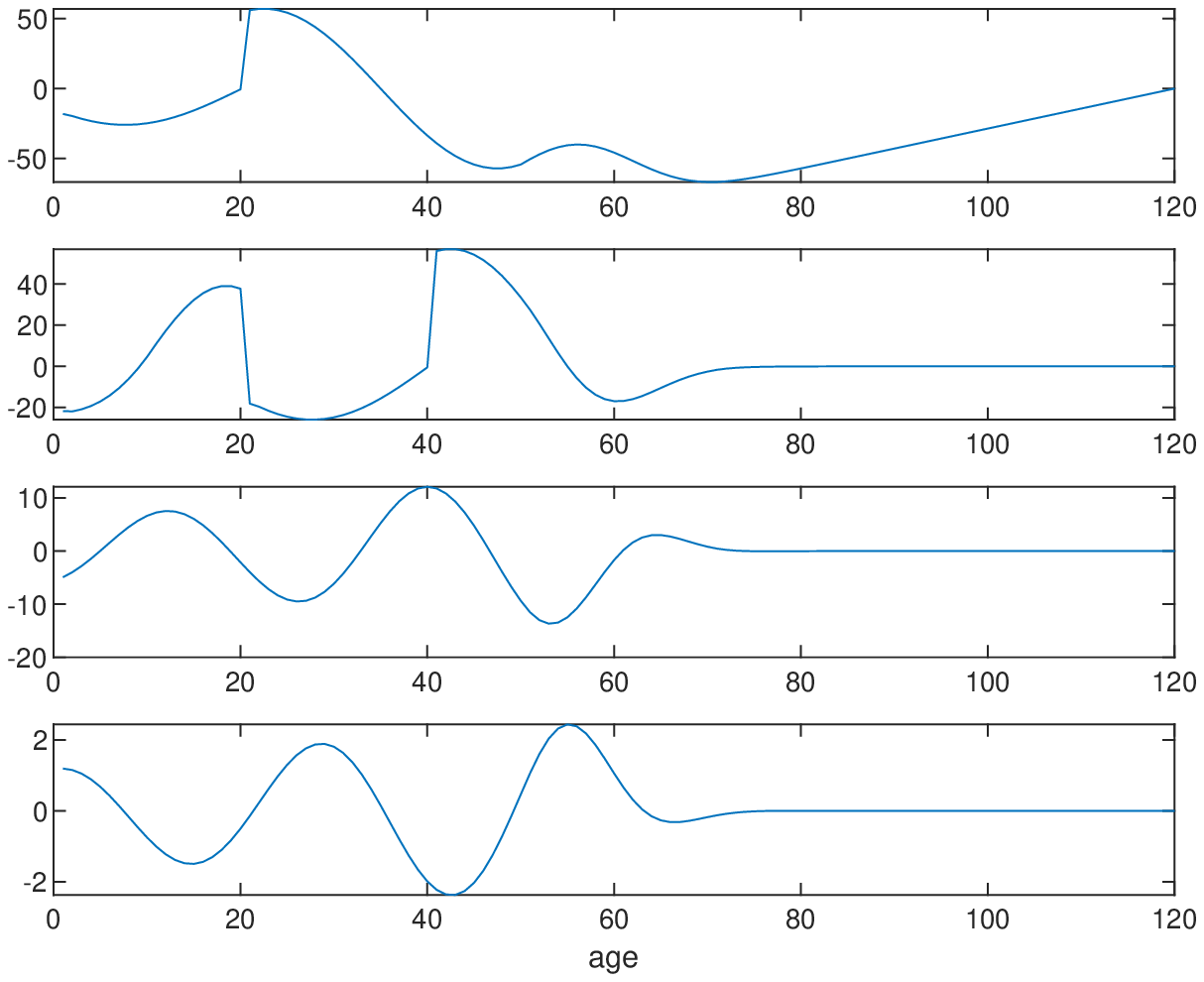}}
\caption{(a). With the mixed birth rates in Figure \ref{mix} and a general initial condition, from the top to the bottom is the temporal evolution of the population distribution: $p(0,a)$; $p(20,a)$; $p(100,a)$; $p(200,a)$. The total birth rate $\al = \al_n = 1.0001$ for $n=50$. (b). The temporal evolution of the total population $P(t)$. (c). From the top to the bottom is the temporal evolution of the population distribution difference: $p(20,a)-p(0,a)$; 
$p(40,a)-p(20,a)$; $p(120,a)-p(100,a)$; $p(220,a)-p(200,a)$. }
\label{GI}
\end{figure}

Replacing the initial condition with general initial conditions does not change the nature of the dynamics either (see 
Figure \ref{GI}). When $\al = 1$, the  weighted averaging in the birth dynamics keeps reducing the maximum and increasing 
the minimum of the population distribution  in the region $a \leq 40$. This leads to the constant asymptotics of the 
population distribution before the sharp transition. 

\begin{figure}[ht] 
\centering
\subfigure{\includegraphics[width=2.3in,height=2.3in]{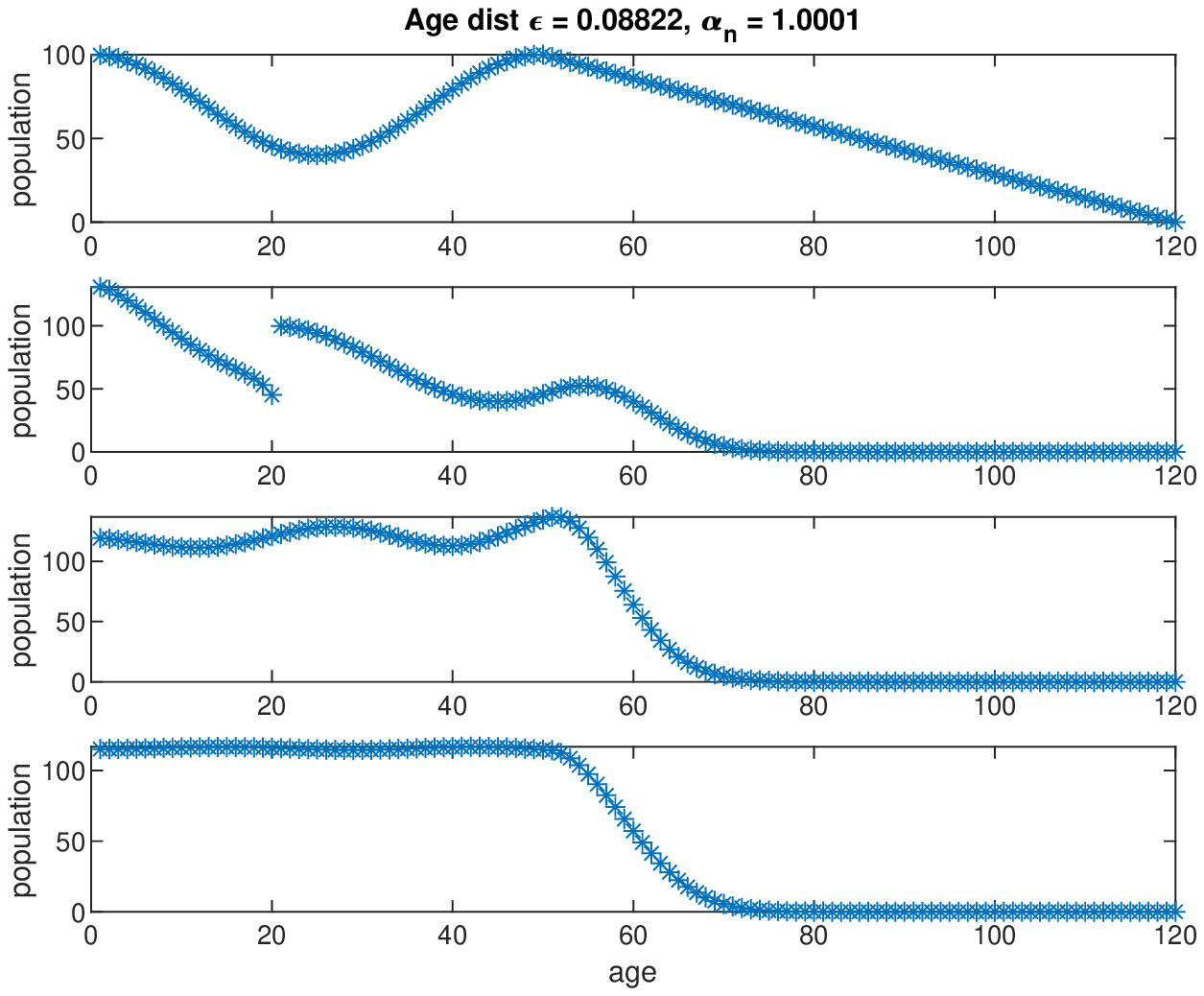}}
\subfigure{\includegraphics[width=2.3in,height=2.3in]{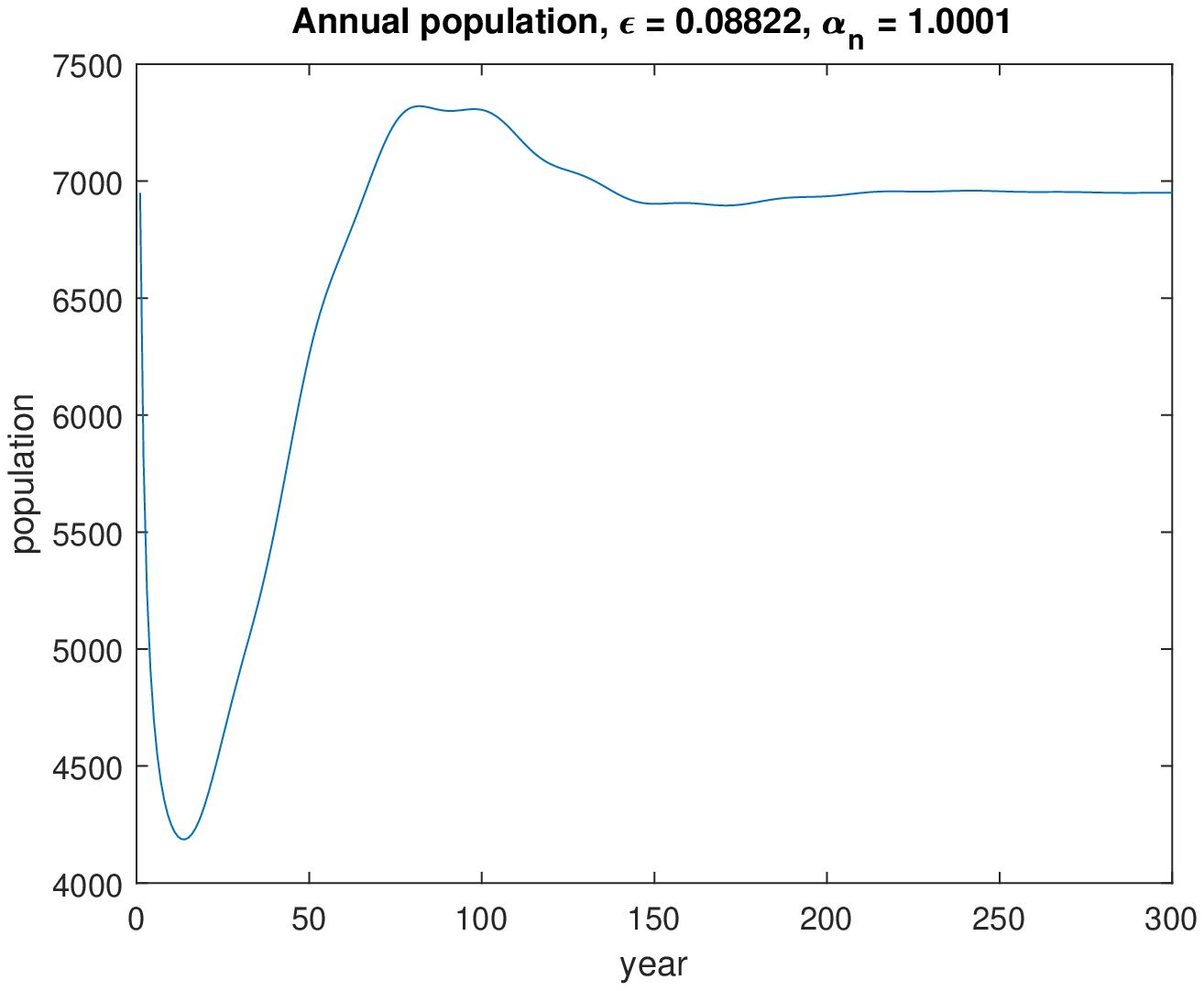}}
\subfigure{\includegraphics[width=2.3in,height=2.3in]{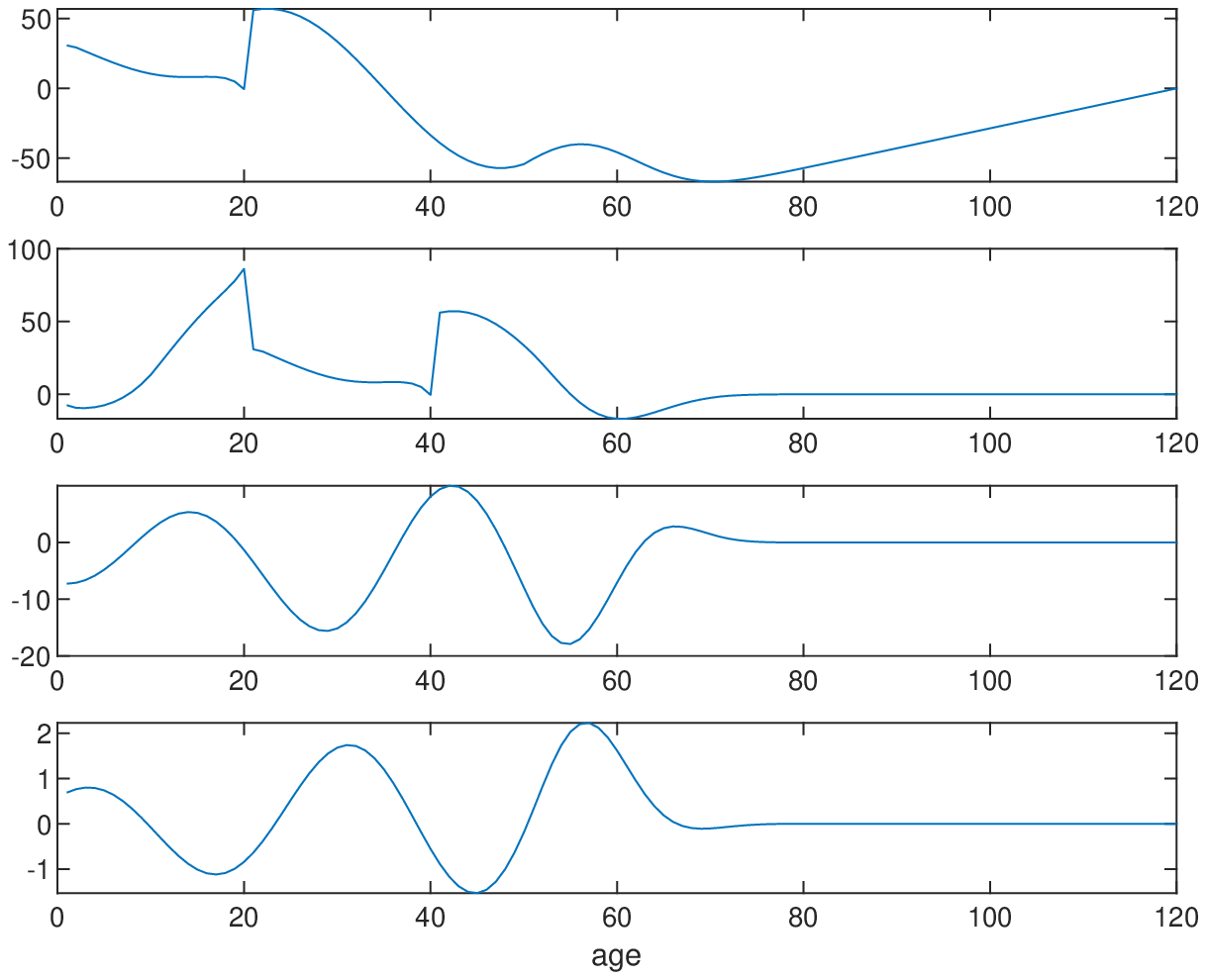}}
\caption{(a). With the nonlinear birth rate (\ref{NLBR}) where $\al (t,b)$ is given in Figure \ref{mix},
from the top to the bottom is the temporal evolution of the population distribution: $p(0,a)$; $p(20,a)$; $p(100,a)$; $p(200,a)$. The total birth rate $\al = \al_n = 1.0001$ for $n=50$. (b). The temporal evolution of the total population $P(t)$. (c). From the top to the bottom is the temporal evolution of the population distribution difference: $p(20,a)-p(0,a)$; 
$p(40,a)-p(20,a)$; $p(120,a)-p(100,a)$; $p(220,a)-p(200,a)$.}
\label{NL1}
\end{figure}

\begin{figure}[ht] 
\centering
\subfigure{\includegraphics[width=2.3in,height=2.3in]{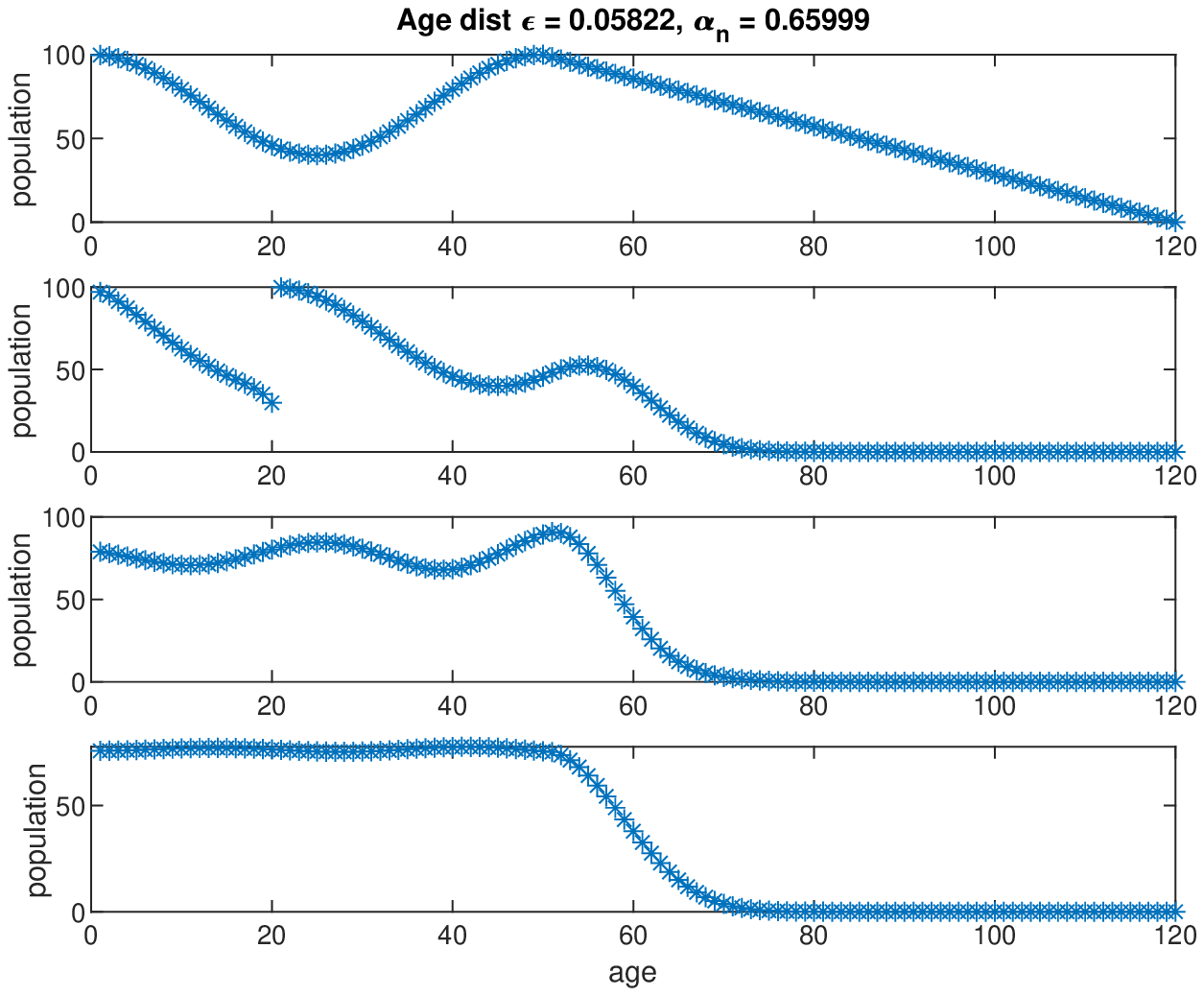}}
\subfigure{\includegraphics[width=2.3in,height=2.3in]{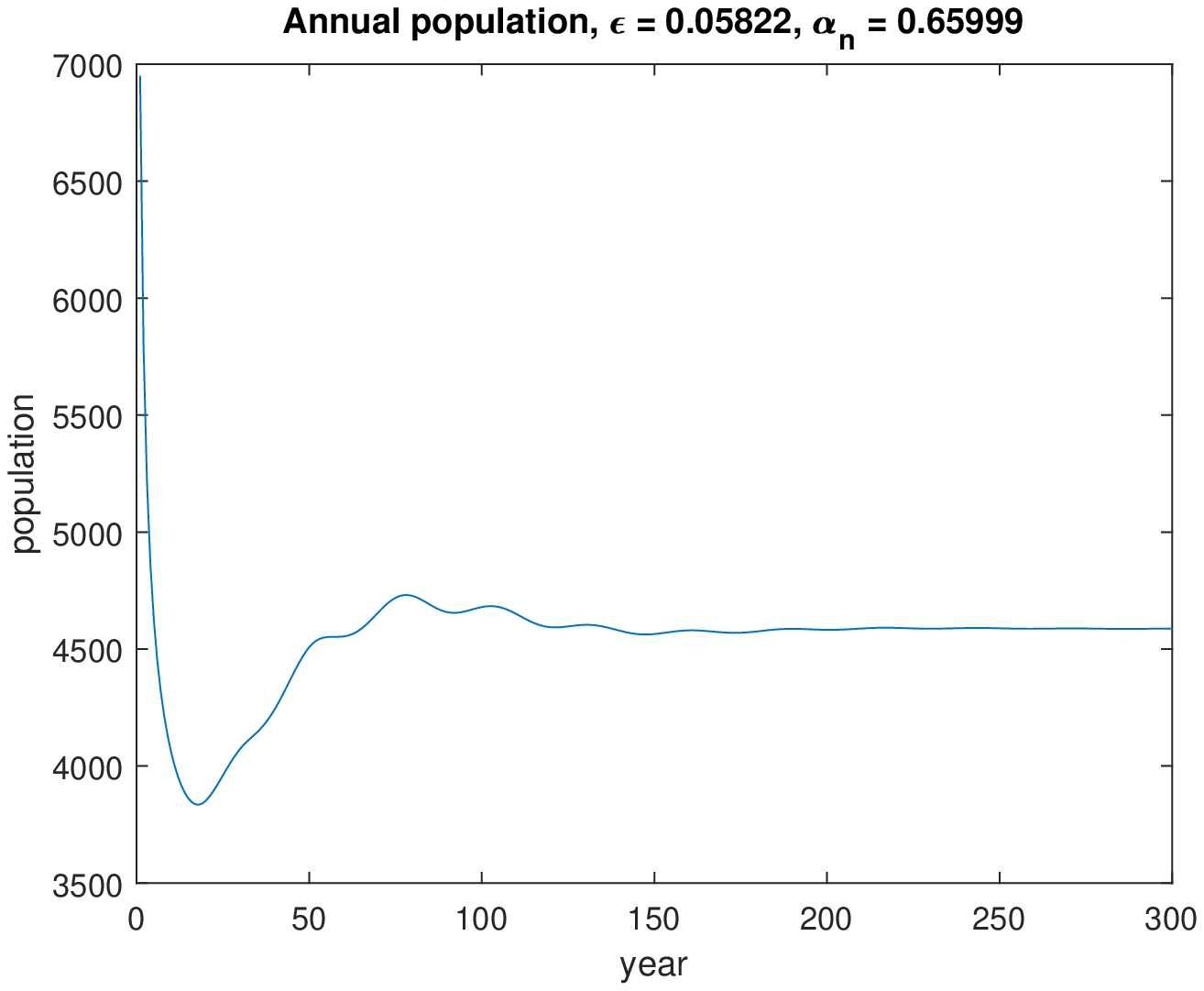}}
\subfigure{\includegraphics[width=2.3in,height=2.3in]{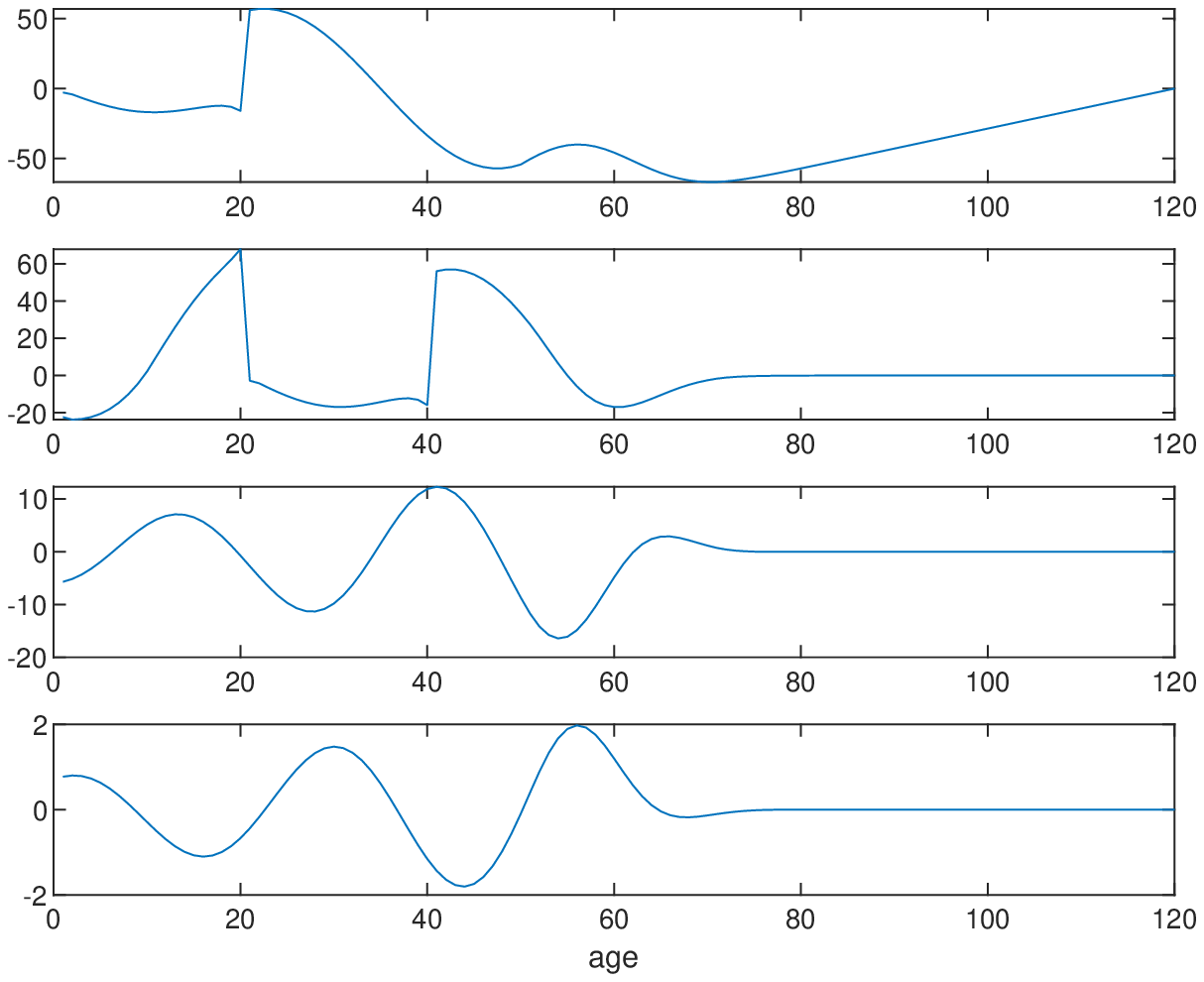}}
\caption{(a). With the nonlinear birth rate (\ref{NLBR}) where $\al (t,b)$ is given in Figure \ref{mix}, from the top to the bottom is the temporal evolution of the population distribution: $p(0,a)$; $p(20,a)$; $p(100,a)$; $p(200,a)$. The total birth rate $\al = \al_n = 0.65999$ for $n=50$. (b). The temporal evolution of the total population $P(t)$. (c). From the top to the bottom is the temporal evolution of the population distribution difference: $p(20,a)-p(0,a)$; 
$p(40,a)-p(20,a)$; $p(120,a)-p(100,a)$; $p(220,a)-p(200,a)$.}
\label{NL2}
\end{figure}

\begin{figure}[ht] 
\centering
\subfigure{\includegraphics[width=2.3in,height=2.3in]{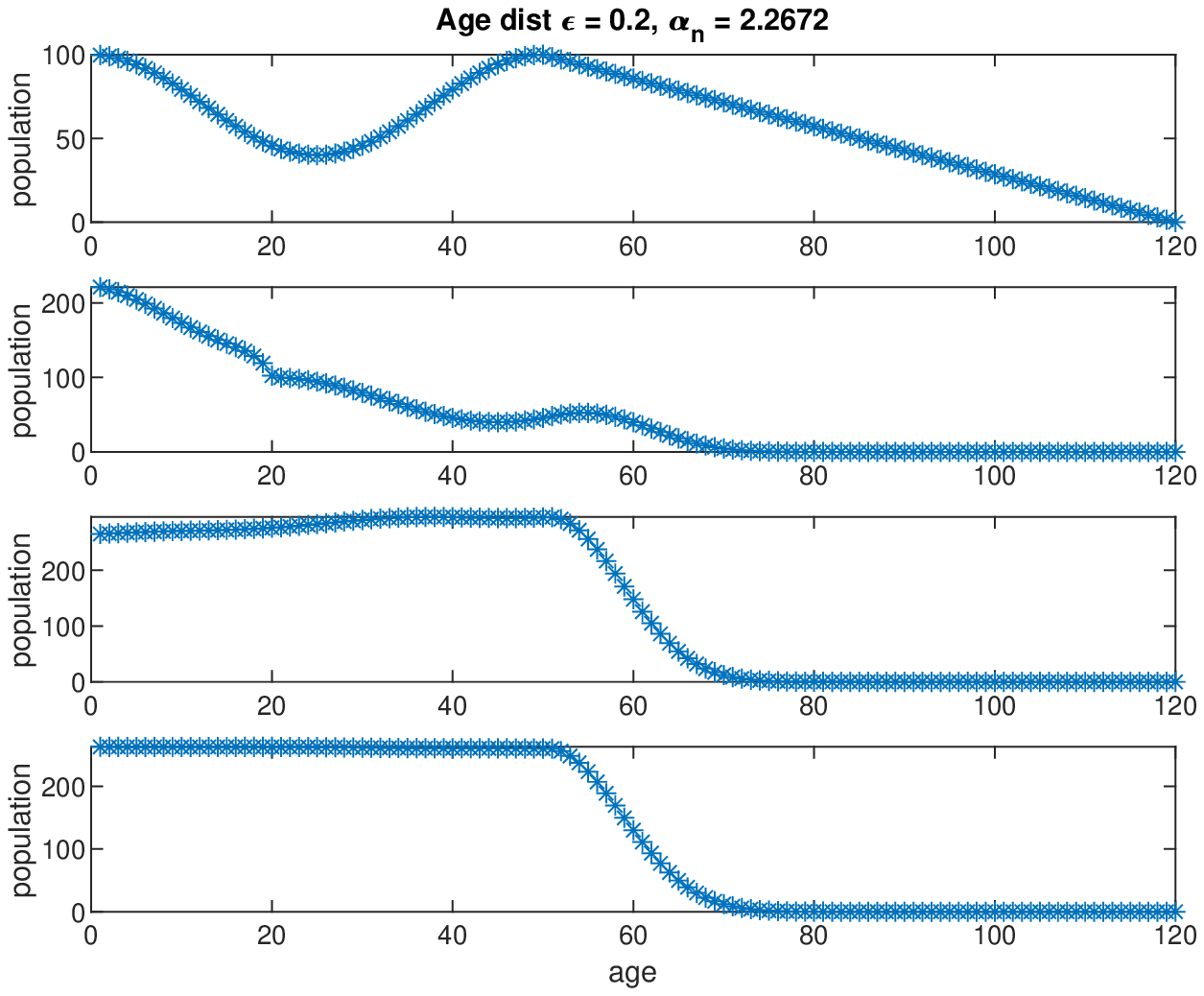}}
\subfigure{\includegraphics[width=2.3in,height=2.3in]{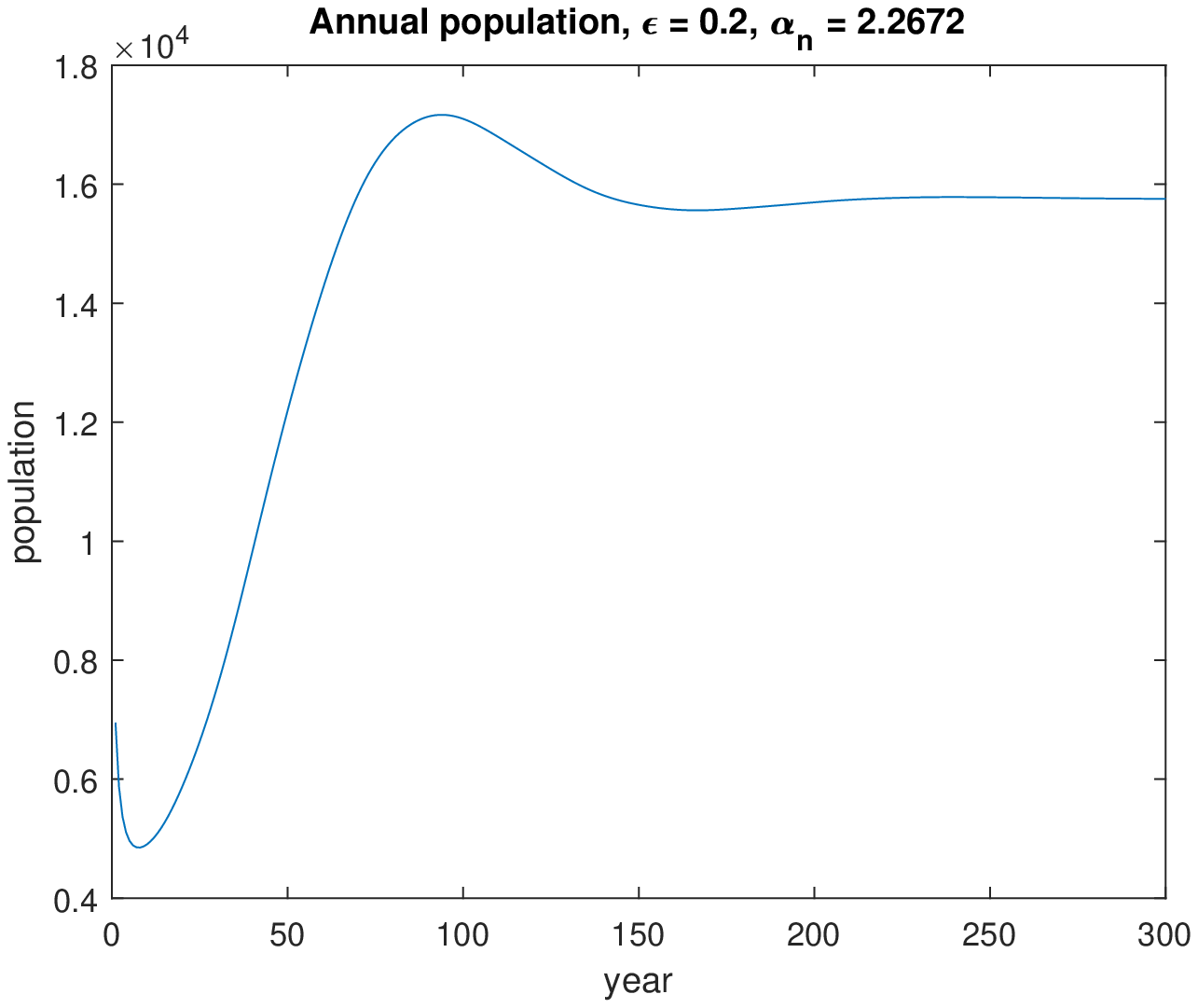}}
\subfigure{\includegraphics[width=2.3in,height=2.3in]{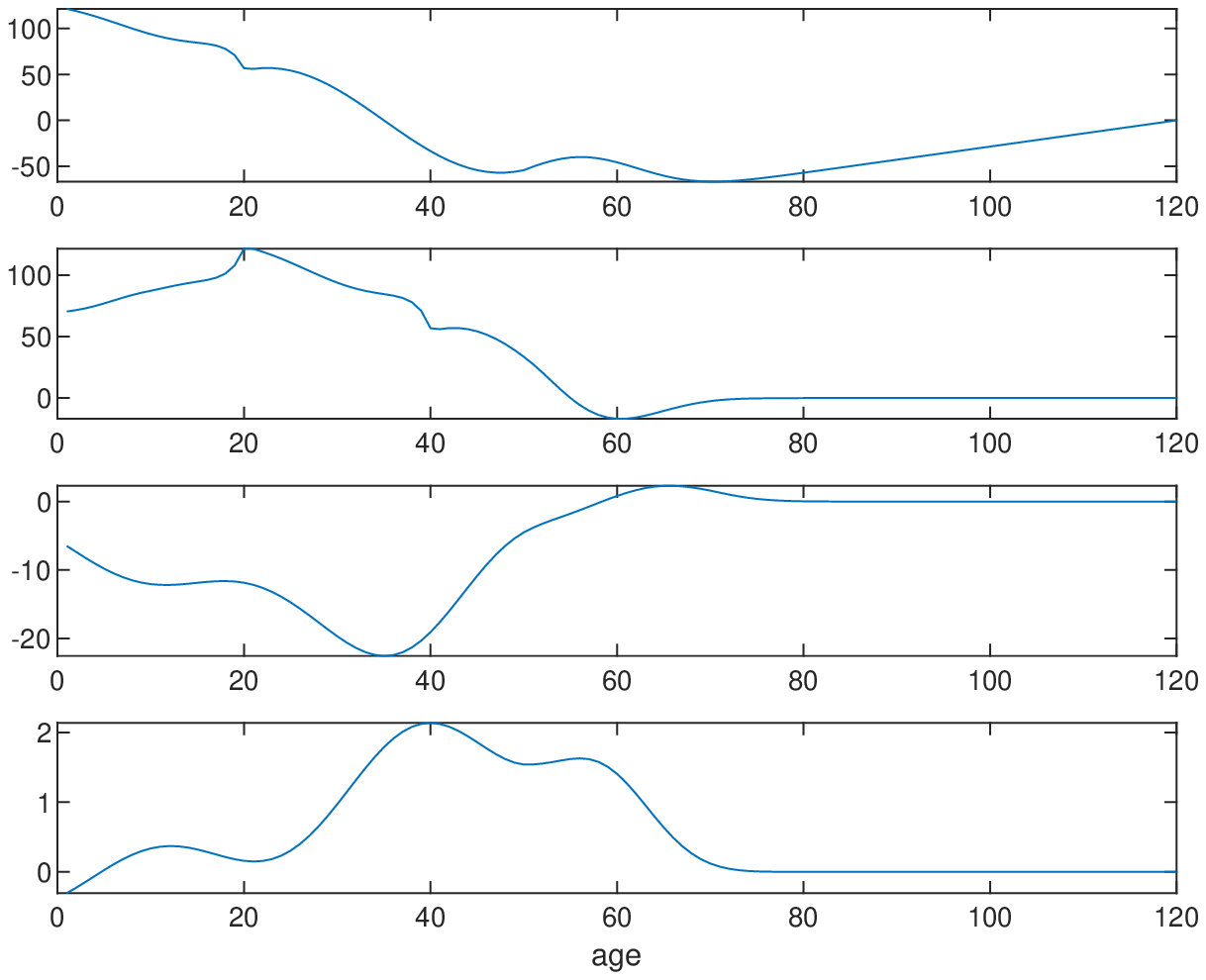}}
\caption{(a). With the nonlinear birth rate (\ref{NLBR}) where $\al (t,b)$ is given in Figure \ref{mix}, from the top to the bottom is the temporal evolution of the population distribution: $p(0,a)$; $p(20,a)$; $p(100,a)$; $p(200,a)$. The total birth rate $\al = \al_n = 2.2672$ for $n=50$. (b). The temporal evolution of the total population $P(t)$. (c). From the top to the bottom is the temporal evolution of the population distribution difference: $p(20,a)-p(0,a)$; 
$p(40,a)-p(20,a)$; $p(120,a)-p(100,a)$; $p(220,a)-p(200,a)$.}
\label{NL3}
\end{figure}

Replacing the birth rates $\al (t,b)$ in (\ref{PAT2}) with 
\begin{equation}
\al (t,b) \frac{P(0)}{P(t)}, \label{NLBR}
\end{equation}
we have the nonlinear birth rates that we are interested in. The idea is that when the total population $P(t)$ increases,
the birth rate decreases. The total birth rate is replaced by $\al \frac{P(0)}{P(t)}$. When $\al \frac{P(0)}{P(t)} >1$, 
the total population increases (as in Figure \ref{A132}), thus $\al \frac{P(0)}{P(t)}$ will decrease. When 
$\al \frac{P(0)}{P(t)} <1$, the total population decreases (as in Figure \ref{A77}), thus $\al \frac{P(0)}{P(t)}$ will increase. 
This causes that $\al \frac{P(0)}{P(t)}$ approaches $1$ as $t \ra +\infty$. Thus the population distribution $p(t,a)$ always approaches an asymptotically steady state with the total population
\begin{equation}
P(t) = \al P(0),
\label{nas}
\end{equation}
given by $\al P(0) /P(t) = 1$ (see Figures \ref{NL1} - \ref{NL3}). 

\section{Conclusion}

A population-age-time (PAT) model is introduced that models the temporal evolution of population distribution in age. The 
model is focused on the effects of birth rate and death rate distribution in age, ignoring the effects of immigration/emigration and migration. To our surprise, the qualitative nature of the population distribution is very robust with respect to different forms 
of initial conditions, birth rate distributions, and death rate distributions. This indicates that the population distributions often take 
certain universal asymptotic shape - a kink, which indeed agrees with typical population distribution in reality (such as the population distribution in US). When the number of children born per woman is $2$, the population distribution approaches an asymptotically steady state of a kink shape; thus the total population approaches a constant.
When the number of children born per woman is greater than  $2$, the total population increases without bound; and 
When the number of children born per woman is less than  $2$, the total population decreases to zero.

\end{document}